\documentclass{aastex631}
\usepackage{amsmath}
\usepackage{soul}
\usepackage{graphicx}
\usepackage{subfigure}
\usepackage{fourier}
\usepackage{arcs}
\usepackage{url}
\usepackage{color}
\usepackage{amsbsy}
\usepackage{mathrsfs}
\usepackage{bm}
\usepackage{booktabs}
\usepackage{multirow}
\usepackage{url}
\usepackage{ulem}
\usepackage{makecell}
\usepackage{tikz}
\begin{document}

\title{SIP-IFVM: An observation-based magnetohydrodynamic model of coronal mass ejection}

\author[0000-0002-4217-6990]{Hao P. Wang}
\affiliation{Centre for Mathematical Plasma-Astrophysics, Department of Mathematics, KU Leuven, Celestijnenlaan 200B, 3001 Leuven, Belgium; \url{haopeng.wang1@kuleuven.be}; \url{jinhan.guo@kuleuven.be}; \url{Stefaan.Poedts@kuleuven.be}}

\author[0000-0002-4205-5566]{Jin H. Guo}
\affiliation{Centre for Mathematical Plasma-Astrophysics, Department of Mathematics, KU Leuven, Celestijnenlaan 200B, 3001 Leuven, Belgium}
\affiliation{School of Astronomy and Space Science and Key Laboratory of Modern Astronomy and Astrophysics, Nanjing University, Nanjing 210023, China}

\author[0000-0002-1743-0651]{S. Poedts}
\affiliation{Centre for Mathematical Plasma-Astrophysics, Department of Mathematics, KU Leuven, Celestijnenlaan 200B, 3001 Leuven, Belgium}
\affiliation{Institute of Physics, University of Maria Curie-Skłodowska, ul. Radziszewskiego 10, 20-031 Lublin, Poland}

\author[0000-0003-4017-215X]{A. Lani}
\affiliation{Centre for Mathematical Plasma-Astrophysics, Department of Mathematics, KU Leuven, Celestijnenlaan 200B, 3001 Leuven, Belgium}
\affiliation{Von Karman Institute For Fluid Dynamics, Waterloosesteenweg 72, 1640 Sint-Genesius-Rode, Brussels, Belgium}

\author[0000-0002-4014-1815]{L. Linan}
\affiliation{Centre for Mathematical Plasma-Astrophysics, Department of Mathematics, KU Leuven, Celestijnenlaan 200B, 3001 Leuven, Belgium}

\author[0000-0002-1986-4496]{T. Baratashvili}
\affiliation{Centre for Mathematical Plasma-Astrophysics, Department of Mathematics, KU Leuven, Celestijnenlaan 200B, 3001 Leuven, Belgium}

\author[0000-0003-4716-2958]{Li P. Yang}
\affiliation{SIGMA Weather Group, State Key Laboratory of Solar Activity and Space Weather, National Space Science Center, Chinese Academy of Sciences, Beijing 100190, People's Republic of China} 

\author[0000-0003-4616-947X]{H.-J. Jeong}
\affiliation{Centre for Mathematical Plasma-Astrophysics, Department of Mathematics, KU Leuven, Celestijnenlaan 200B, 3001 Leuven, Belgium}
\affiliation{School of Space Research, Kyung Hee University, Yongin, 17104, Republic of Korea}

\author[0000-0001-8495-9179]{Wen W. Wei}
\affiliation{Space Sciences Laboratory, University of California, Berkeley, CA 94720, USA}

\author{Cai X. Li}
\affiliation{Shenzhen Key Laboratory of Ultraintense Laser and Advanced Material Technology, Center for Intense Laser Application Technology, and College of Engineering Physics, Shenzhen Technology University, Shenzhen 518118, China}

\author{Y. Yang}
\affiliation{School of Mathematical Sciences, Ministry of Education Key Laboratory for NSLSCS, Nanjing Normal University, Nanjing 210023, People's Republic of China}

\author{Yu C. Li}
\affiliation{Centre for Mathematical Plasma-Astrophysics, Department of Mathematics, KU Leuven, Celestijnenlaan 200B, 3001 Leuven, Belgium}
\affiliation{SIGMA Weather Group, State Key Laboratory of Space Weather, National Space Science Center, Chinese Academy of Sciences, Beijing 100190, People's
Republic of China}
\affiliation{College of Earth and Planetary Sciences, University of Chinese Academy of Sciences, Beijing 100049, People's Republic of China}

\author{Hao. Wu}
\affiliation{School of Astronomy and Space Science and Key Laboratory of Modern Astronomy and Astrophysics, Nanjing University, Nanjing 210023, China}

\author{Yang. Guo}
\affiliation{School of Astronomy and Space Science and Key Laboratory of Modern Astronomy and Astrophysics, Nanjing University, Nanjing 210023, China}

\author[0000-0003-3364-9183]{B. Schmieder}
\affiliation{Centre for Mathematical Plasma-Astrophysics, Department of Mathematics, KU Leuven, Celestijnenlaan 200B, 3001 Leuven, Belgium}
\affiliation{Observatoire de Paris, LIRA, UMR8254 (CNRS), F-92195 Meudon Principal Cedex, France}
\affiliation{SUPA, School of Physics $\&$ Astronomy, University of Glasgow, Glasgow G12 8QQ, UK}

\begin{abstract}

Coronal mass ejections (CMEs) can cause severe space weather disturbances, making it essential to develop an efficient and numerically stable magnetohydrodynamic (MHD) model capable of accurately and promptly simulating realistic CME evolution.
However, achieving a balance between computational efficiency, accuracy, and numerical stability in CME simulations, particularly in the sub-Alfv{'e}nic coronal region, remains a significant challenge.
This paper aims to address the challenge by integrating observational data and developing advanced numerical algorithms, focusing on reproducing large-scale CME evolutions that are consistent with observations in the coronal region.
Based on the recently developed fully implicit thermodynamic MHD coronal model \citep{wang2025sipifvmtimeevolvingcoronalmodel}, we further use an observation-based Regularised Biot-Savart Laws (RBSL) flux rope to trigger a CME event during Carrington rotation (CR) 2111. Additionally, we improve the temporal accuracy using a 2nd-order accurate Explicit Singly Diagonally Implicit Runge-Kutta (ESDIRK2) method, with the intermediate stage solutions computed by the 2nd-order accurate Backward Differentiation Formula (BDF2) pseudo-time marching method. To enhance the numerical stability of ESDIRK2, we apply approximate linearisation in the implicitly solved intermediate stages. Furthermore, we adjust the time-evolving magnetic field $\mathbf{B}_1$ to zero at the end of each physical time step to further validate the extended magnetic field decomposition approach proposed by \cite{wang2025sipifvmtimeevolvingcoronalmodel}.
It is noticed that the model successfully reproduces the CME evolution consistent with white-light coronagraph observations, enables faster-than-real-time CME propagation simulations from solar surface to 0.1 AU using only a few dozen CPU cores on approximately 1 million grid cells, and remains numerically stable in CME simulations involving low-$\beta$ regions.
The simulation results show that this novel MHD coronal model, combined with an observation-based magnetic flux rope, is sufficiently numerically stable and computationally efficient to reproduce real CME events propagating through the sub-Alfv{\'e}nic coronal region. Thus, the observation-based CME model is well suited for practical applications in daily space weather forecasting.

\end{abstract}

\section{Introduction}\label{sec:intro}
Coronal mass ejections (CMEs) are expulsion of magnetic plasma from the corona into the interplanetary medium \citep{Chen2025,Linan2023,Linker2024,Torok_2018,VERBEKE20235243}, and are manifested as bright features in the solar corona observed in white light with coronagraphs \citep{Manchester_IV2004,Schmieder2015}. They are among the most energetic phenomena in the solar system \citep{Gou2023}, leading to solar energetic particle events \citep{Desai2016,Vlahos2019} and geomagnetic storms \citep{Gonzalez,Lugaz2016}, and are dominant contributors to adverse space weather when Earth-directed \citep{WangWS2019}. 
Given that severe space weather can affect the performance and reliability of both space- and ground-based technological systems and pose risks to human health, developing a physics-based and MHD model of CMEs that can timely and accurately reproduce their evolution and propagation can not only promote scientific research work but also help mitigate societal risks.

Currently, one of the main challenges in CME modelling is balancing computational efficiency, accuracy, and numerical stability for simulations in the sub-Alfv{\'e}nic coronal region, corresponding to the early stages of CME propagation and shock evolution from solar surface to below about 0.1 AU \citep{Vourlidas2019}. Within this region, especially near the solar surface, plasma $\beta$ (the ratio of thermal pressure to magnetic pressure) can sometimes drop as low as $10^{-4}$ \citep{Bourdin2017}. In such low-$\beta$ regions, MHD codes often suffer from extremely low computational efficiency due to the extremely small time steps determined by the Courant–Friedrichs–Lewy (CFL) stability condition \citep{Feng_2021,Wang_2022,Wang2022_CJG,WangSubmitted}, and struggle with severe numerical stability issues caused by appearance of nonphysical negative thermal pressures, which arise from magnetic energy discretization errors when thermal pressure is derived from the total energy density \citep{W_SubmittedCOCONUT,wang2025sipifvmtimeevolvingcoronalmodel}. Consequently, MHD coronal models, by which the CME evolution and propagation are calculated, become the most complex and computationally intensive component in the Sun-to-Earth model chain \citep{WangSubmitted,W_SubmittedCOCONUT}. However, timely and effective space weather forecasting requires Sun-to-Earth model chains \cite[e.g.,][]{Feng_2011Chinese,Feng_2013Chinese,Pomoell2018020,Poedts_2020,Hayashi_2021} to provide reliable predictions hours to days in advance \cite[e.g.][]{BAKER19987,Feng_2013Chinese,Koskinen2017,Feng2020book}, highlighting the need to develop sufficiently efficient and reliable CME models in the coronal region. The book \cite{Feng2020book} presented a thorough survey of various coronal and interplanetary MHD models  at the time it was written.

   Implicit methods that allow for a considerable time step exceeding the CFL stability condition have enabled both quasi-steady-state MHD coronal models constrained by a static magnetogram \citep{Perri2018SimulationsOS,WANG201967,Feng_2021,Perri_2022,Wang_2022,Wang2022_CJG,Kuzma_2023,Perri_2023,WangSubmitted} and time-evolving coronal models driven by a series of time-evolving magnetograms \citep{W_SubmittedCOCONUT,wang2025COCONUTMayEvent,wang2025sipifvmtimeevolvingcoronalmodel} to achieve the desired speedup. COolfluid COroNal UnsTructured (COCONUT), a novel implicit MHD solar corona model based on Computational Object-Oriented Libraries for Fluid Dynamics (COOLFluiD) \citep{kimpe2,lani1,lani13}\footnote{\url{https://github.com/andrealani/COOLFluiD/wiki}}, and the Solar-Interplanetary Phenomena-Implicit Finite Volume Method (SIP-IFVM) coronal models \citep{WangSubmitted,wang2025sipifvmtimeevolvingcoronalmodel} are two prominent examples of the successful application of implicit strategies in coronal modelling. It demonstrates that the implicit temporal integration approach can reduce the wall-clock time of quasi-steady-state MHD coronal models from dozens of hours to less than one hour and can achieve a faster-than-real-time performance for the time-evolving MHD coronal simulations with only a few dozens of CPU cores. In this paper, we implement the implicit method in a CME simulation calculated by solving thermodynamical MHD equations, and adopt an approximate large time step to improve computational efficiency.

   In unsteady-state simulations, extra measures are required to guarantee a desired temporal accuracy when approximate large time steps are adopted in implicit methods. During time-dependent simulations, COCONUT employs the BDF2 scheme for temporal integration and performs Newton iterations within each physical time step to update the intermediate solution states \citep{W_SubmittedCOCONUT}. Alternatively, SIP-IFVM utilised the backward Euler method (BDF1) for temporal integration and improves temporal accuracy using a pseudo-time marching method \cite[e.g.,][]{BIJL2002313,Lingquan2019,LUO2001137,SITARAMAN2013364}, which introduces a pseudo time $\tau$ at each physical time step and solves a steady-state problem in $\tau$ using the backward Euler method. Given that the backward Euler method is only 1st-order accurate in time, while SIP-IFVM is 2nd-order spatial accurate, we further improve the temporal integration in this paper to achieve 2nd-order accuracy as well to reduce the excessive numerical dissipation. Since explicit singly diagonally implicit Runge-Kutta (ESDIRK) method can be both A-stable and L-stable regardless of its order\citep{BIJL2002313,KENNEDY2003139,KENNEDY2019221,NOVENTA2020104529}, we adopt a 2nd-order accurate three-stage ESDIRK2 method \citep{KennedyCarpenter2016,WangLai2020,WANG2020109581}, where the first stage is achieved explicitly, and the 2nd and 3rd stages are solved implicitly. Similarly to \cite{WangSubmitted}, we still employ a pseudo-time marching method to calculate the solutions in these two stages. However, instead of the backward Euler method, we adopt the 2nd-order accurate BDF2 scheme \citep{Nived2023} to more accurately advance the intermediate solutions to a steady state in $\tau$ within limited iterations during these two stages.

   The numerical stability of MHD coronal models has also been significantly improved by various methods. The positivity-preserving (PP) methods in \cite{wang2025COCONUTMayEvent} enable the time-evolving COCONUT to deal with coronal evolutions during solar maximum with a magnetic field as strong as about 30 Gauss. Some specific measures, such as switching to a magneto-frictional model when driving coronal evolution below 1.15$\;R_s$ \citep{Hoeksema2020,Hayashi_2021}, and employing a solution-preserving method that artificially increases plasma density to broaden the transition region \citep{Lionello_2008,Mikic_2013,MIKIC2018NatA,Mok_2005}, have also been adopted to enhance the numerical stability of MHD coronal models. Additionally, \cite{wang2025sipifvmtimeevolvingcoronalmodel} proposed the extended magnetic field decomposition strategy, which enables the SIP-IFVM coronal model to perform well in time-evolving coronal simulations with $\beta$ being as small as around $10^{-3}$. In addition to a static potential field and a time-evolving magnetic field component $\mathbf{B}_1$, it introduces a temporally piecewise-constant variable to accommodate part of the non-potential field, helping to remain $\mathbf{B}_1$ sufficiently small throughout the simulation and thereby improve the numerical stability of the time-evolving coronal model in dealing with low-$\beta$ issues. In \cite{wang2025sipifvmtimeevolvingcoronalmodel}, $\mathbf{B}_1$ is reset to zero when a specific critical condition is reached, and the time-evolving coronal simulation does not involve strong electric currents as large as those appearing near the flux rope in CME simulations. Therefore, its effectiveness in CME simulations remains unclear yet. In this paper, we implement the extended magnetic field decomposition strategy to CME simulations and further validate its effectiveness by resetting $\mathbf{B}_1$ to zero after each solution update at every physical time step.

   With the coronal background prepared, we still need to trigger a CME event to further simulate its evolution together with the surrounding plasma environment. In recent decades, many researchers have used various magnetic flux-rope models to trigger CME events in MHD coronal models. Such as the Gibson–Low (GL) flux rope model \citep{Gibson_1998} adopted in AWSoM \citep{Jin_2017}, analytical modified Titov-D\'{e}moulin circular \citep{Titov_2014} flux ropes (TDm) and RBSL \citep{Titov_2018} flux ropes implemented in COCONUT \citep{guo2023,Linan_2023,Linan_2025}, SIP-IFVM \cite{WangSubmitted}, MAS \citep{Linker2024}, PLUTO \citep{Regnault_2023} and MPI-AMRVAC \citep{Guo2019,Keppens2023} codes, and  spheroid flux rope model performed in SIP-CESE MHD mode \citep{Yang_2021}. A more detailed description of these models is available in \cite{WangSubmitted}. Among these simulations, the MAS coronal model determines the CME source regions by constructing the magnetic flux rope along the elongated and curved polarity inversion line (PIL) \citep{Titov_2018,Linker2024}, the input magnetogram and observed CME speed (from SOHO/LASCO and/or STEREO/COR observations) are used to determine the initial GL flux rope used in AWSoM \citep{Jin_2017} to match CME observations. In this paper, we further initialize an RBSL flux rope by specifying its initial position, axis, and magnetic flux based on observational data, demonstrating the extent to which the observation-based CME simulation can reproduce the observed CME evolution within 0.1 AU.

   In \cite{WangSubmitted}, we have made a faster-than-real-time CME propagation simulation from the solar surface to 0.1 AU. It adopted backward Euler temporal integration to improve computational efficiency by an approximate large time step, used the improved parallel LU-SGS matrix solver \citep{Feng_2021,Otero2015,OteroAcc2015,Wang_2022}, which enhances the diagonal dominance of the Jacobian matrix, to solve the linearised system, and employed the pseudo-time marching method to improve the temporal accuracy. Although it balances computational efficiency and temporal accuracy well, several issues remain for further improvement. It is a virtual CME event triggered by an analytical RBSL flux rope in a relatively simple solar minimum coronal background, making it less convincing than a realistic CME event simulation in more complex solar maximum coronal background. Radiative loss terms are missed, and the adiabatic index $\gamma$ equals 1.05, closely approximating an isothermal process ($\gamma = 1$). Additionally, a limiter is still required for reconstruction formula of the magnetic field to suppress spatial oscillations during the CME simulation, which can cause degradation in precision and divergence-free constraint of the magnetic field. These issues mentioned above are addressed in this paper. 

   Based on the above considerations, this paper is organised as follows. In Section~\ref{sec:Description of the novel MHD Model}, we present the numerical formulation and provide a detailed description of the ESDIRK2 temporal integration method, including the approximate linearisation adopted in the intermediate stages. This section also describes the implementation of observation-based CME simulations using the implicit MHD coronal model. In Section~\ref{sec:Numerical Results}, we present the simulation results, including the comparison between the simulation results calculated by the extended and conventional magnetic field decomposition method. Evolution of open-field regions, selected magnetic field lines, two-dimensional (2D) timing diagrams of physical parameters along the heliocentric distance during the CME simulation, and comparisons between synthetic simulated pB images and coronagraph observations are also presented. In Section~\ref{sec:conclusion}, we summarise the key features of the efficient observation-based CME propagation modelling approach and provide concluding remarks.

\section{Numerical algorithm}\label{sec:Description of the novel MHD Model}
\subsection{Governing equations and initial value setup}\label{The governing equations} 
We solve the extended decomposed MHD equations, in which the magnetic field $\mathbf{B}$ is divided into a time-independent potential field $\mathbf{B}_{00}$, a temporally piecewise constant field $\mathbf{B}_{01}$, and a time-dependent field $\mathbf{B}_1$ \citep{wang2025sipifvmtimeevolvingcoronalmodel}, to conduct both the quasi-steady-state background solar coronal simulation and the following CME evolution and propagation simulation triggered by an observation-based flux rope. However, instead of waiting for $\mathbf{B}_1$ to reach a specific threshold, we adjust $\mathbf{B}_{01}$ to accommodate $\mathbf{B}_1$ and then reset $\mathbf{B}_1$ to zero at the end of each physical time step. The governing equations are the same as those in \cite{wang2025sipifvmtimeevolvingcoronalmodel} and are described in the following compact form:
\begin{equation}\label{MHDinsolarwind}
\frac{\partial \mathbf{U}}{\partial t}+\nabla \cdot \mathbf{F}\left(\mathbf{U}\right)=\mathbf{S}\left(\mathbf{U},\nabla \mathbf{U}\right).\
\end{equation}
   Here, $\mathbf{U}=\left(\rho, \rho \mathbf{v}, E_1, \mathbf{B}_1\right)^T$ is the conservative variable vector with $E_1=\frac{p}{\gamma-1}+\frac{1}{2}\rho \mathbf{v}^2+\frac{1}{2}\mathbf{B}_1^2$. $\rho$, $p$ and $\mathbf{v}=\left(u, v, w\right)^{T}$ are the plasma density, thermal pressure and velocity, respectively. The adiabatic index is $\gamma=\frac{5}{3}$. $\nabla \mathbf{U}$ means the derivative of $\mathbf{U}$, $\mathbf{F}\left(\mathbf{U}\right)=\left(\mathbf{f}\left(\mathbf{U}\right),\mathbf{g}\left(\mathbf{U}\right),\mathbf{h}\left(\mathbf{U}\right)\right)$ is the inviscid flux vector with $\mathbf{f}\left(\mathbf{U}\right)$, $\mathbf{g}\left(\mathbf{U}\right)$, and $\mathbf{h}\left(\mathbf{U}\right)$ denoting the components in the $x$, $y$, and $z$ directions, and $\mathbf{S}\left(\mathbf{U},\nabla \mathbf{U}\right)$ represents the source term vector containing the Godunov-Powell source terms, the gravitational force, the Coriolis force, and the heating source terms. 
   
   During the background coronal simulation, the radial inner-boundary magnetic field is calculated by a potential field (PF) solver with a 20th-order spherical harmonic expansion. The PF solver is constrained by a GONG-ADAPT (Global Oscillation Network Group–Air Force Data Assimilative Photospheric Flux Transport) magnetogram\footnote{\url{https://gong.nso.edu/adapt/maps/gong/2011/}} from Jun 21, 2011, using the first realization of the 12-member ensemble \citep{LiHuichao2021,Perri_2023}. The initial magnetic field within the computational domain is also determined using this PF solver, and the remaining inner boundary conditions are set following the approach in \cite{wang2025sipifvmtimeevolvingcoronalmodel}. 
   After the background coronal simulation reaches a steady state \citep{Feng_2021,Wang_2022}, the observation-based RBSL flux rope magnetic field, denoted by $\mathbf{B}_{FR}(\mathbf{x})$ with $\mathbf{x}$ being a position vector, is inserted into the corona via a divergence-free radial basic function interpolation \citep{Wang_2022}, following the procedure described in the arXiv v1 version of \cite{WangSubmitted}, to trigger a CME event. 

\begin{figure}[htpb]
\begin{center}
  \vspace*{0.01\textwidth}
    \includegraphics[width=0.8\linewidth,trim=1 1 1 1, clip]{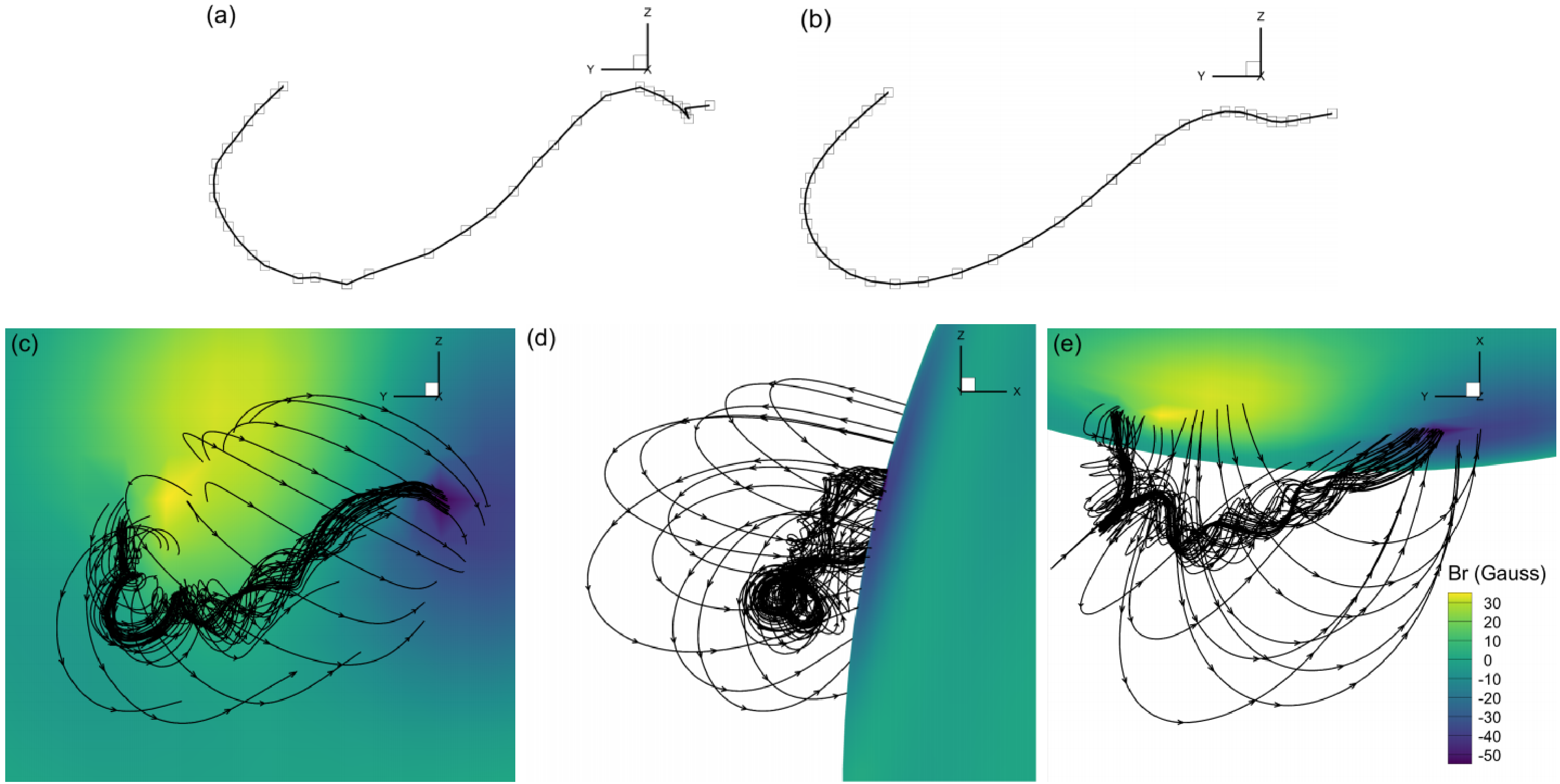}
\end{center}
\caption{Illustration of the RBSL flux rope axis path derived from the original observed (a) and the smoothed (b) sample points. The coordinates of the observed sample points are provided in Appendix~\ref{sec:FRaxispathData}. Panels (c), (d), and (e) show selected magnetic field lines of the flux rope and the background corona, viewed from three mutually perpendicular perspectives. The x-axis is defined as the intersection of the solar equator and the meridional plane passing through the midpoint between the two footpoints of the axis path on the solar surface, pointing to the heliocentric. The z-axis is aligned with the Sun’s north rotational axis, and the y-axis is determined by the right-hand rule. The contour on the solar surface denotes the radial magnetic field strength.}\label{Fluxropepath}
\end{figure}
In Fig.~\ref{Fluxropepath}, we present the original observed flux rope axis path (a) and its smoothed version (b), along with selected magnetic field lines around the flux rope, viewed from three mutually perpendicular perspectives (c, d, e) at the start time of the CME simulation. The radial magnetic field strength in units of Gauss is also displayed by the contour on the solar surface.
The RBSL flux rope magnetic field is calculated as the curl of an axial and an azimuthal vector potential field, generated by a current $I$ and a net flux $F$, respectively \citep{Titov_2018}. In practice, it is fully specified by the magnetic flux $F$, the axis path $C$, and the cross-sectional radius $a$ of $C$. In this paper, derived from observations \citep{Guo2019,Guo_2021}, we set $a = 14~\rm{Mm}$, place two footpoints of $C$ in $\left(\theta, \phi\right) = \left(15.8^\circ~\mathrm{N}, 156.6^\circ\right)$ and $\left(\theta, \phi\right) = \left(15.2^\circ~\mathrm{N}, 171.0^\circ\right)$, and define $C$ by 35 sample points derived from observations. The coordinates of these sample points are provided in Appendix Appendix \ref{sec:FRaxispathData}. To reduce noise in the observed data, a three-point moving average filter is applied to the sample points, resulting in a smoothed axis path that is subsequently used to constrain the RBSL model. We find that $F \approx 1.25 \times 10^{20}~\rm{Mx}$ makes the flux rope field near the footpoints comparable to the background coronal field and performs well in CME simulations. 
In contrast, using $F = 3.7 \times 10^{20}~\rm{Mx}$, the value derived from observations, yields an excessively strong magnetic field relative to the background corona, resulting in extremely high-speed flows and ultimately causing the simulation to crash. 
Therefore, we adopt $F = 1.25 \times 10^{20}~\rm{Mx}$ in this work, which is comparable to the background magnetic flux in this region. 

  \subsection{Solver description}\label{sec:solverdescription} 
  We employ Godunov’s method \citep{EINFELDT1991273,Godunov1959Adifference} to advance the cell-averaged solutions in time by solving a Riemann problem at each cell interface. The governing equations are discretized on a six-component composite mesh \citep{Feng_2010,Feng2020book,Wang_2022,WangSubmitted}, where each component is treated as a low-latitude spherical mesh containing $96 \times 42 \times 42$ cells, a total of approximately one million cells.

  By integrating Eq.~(\ref{MHDinsolarwind}) over cell $i$ and applying Gauss’s theorem to convert the volume integral of the divergence of the flux into a surface integral, we obtain the following spatially discretized integral equations:
\begin{equation}\label{TemporalIntegrationEq}
   V_{i}~\frac{\partial \mathbf{U}_{i}}{\partial t}+\mathbf{R}_{i}=\mathbf{0} \,.
   \end{equation}
   where $\mathbf{R}_i=\sum\limits_{j=1}^{6}\mathbf{F}_{ij}\left(\mathbf{U}_{i,L},\mathbf{U}_{j,R}\right) \cdot \mathbf{n}_{ij}~\Gamma_{ij}-V_{i}~\mathbf{S}_i\left(\mathbf{U},\nabla \mathbf{U}\right)$ means the residual operator over cell $i$. $\mathbf{S}_i\left(\mathbf{U},\nabla \mathbf{U}\right)$ and $V_{i}$ are the cell-average source terms and the volume of cell $i$, and $\Gamma_{ij}$ refers to the interface shared by cell $i$ and cell $j$. $\mathbf{U}_{i,L}$ and $\mathbf{U}_{j,R}$ denote the left and right solution vectors reconstructed on the centroid of $\Gamma_{ij}$ from cell $i$ and cell $j$, respectively, and are reconstructed using the least squares (LSQ) method \citep{BARTH1991,BARTH1993}. To prevent degradation in both accuracy and divergence-free constraint of the magnetic field, we discard the limiter in the reconstruction formulation of the magnetic field. $\mathbf{F}_{ij} \cdot \mathbf{n}_{ij}$ is calculated using the Harten-Lax-van Leer-Riemann solver with a self-adjustable dissipation term \citep{W_SubmittedCOCONUT,wang2025sipifvmtimeevolvingcoronalmodel}. A detailed description of the formulations for $\mathbf{F}_{ij}\left(\mathbf{U}_{i,L},\mathbf{U}_{j,R}\right) \cdot \mathbf{n}_{ij}$ and $\mathbf{S}_i\left(\mathbf{U},\nabla \mathbf{U}\right)$ is available in \cite{Feng_2021} and \cite{Wang_2022,WangSubmitted,wang2025sipifvmtimeevolvingcoronalmodel}.

   We first perform a time-relaxation procedure to solve Eq.~(\ref{TemporalIntegrationEq}) and obtain a quasi-steady background corona, then insert the observation-based RBSL flux rope to trigger a CME event, and finally solve Eq.~(\ref{TemporalIntegrationEq}) by a 2nd-order accurate temporal integration algorithm to simulate the subsequent CME evolution.

   During the background coronal simulation, we use the BDF2 temporal integration method to solve Eq.~(\ref{TemporalIntegrationEq}) as follows:
\begin{equation}\label{implicitbackwardEuler}
   V_{i}\frac{3\mathbf{U}_i^{n+1}-4\mathbf{U}_i^{n}+\mathbf{U}_i^{n-1}}{\Delta t}=-\mathbf{R}^{n+1}_{i}\,.
   \end{equation}
   The superscripts $^{`n'}$ and $^{`n+1'}$ denote the time level, $\mathbf{R}^{n+1}_i=\sum\limits_{j=1}^{6}\mathbf{F}_{ij}\left(\mathbf{U}^{n+1}_{i,L},\mathbf{U}^{n+1}_{j,R}\right) \cdot \mathbf{n}_{ij}~\Gamma_{ij}-V_{i} ~\mathbf{S}_i\left(\mathbf{U}^{n+1},\nabla \mathbf{U}^{n+1}\right)$ represents the residual at the $(n+1)$-th time level, and $\Delta t = t^{n+1} - t^n$ is the time step from the $n$-th to the $(n+1)$-th time level. 
   
   An approximate local temporal linearisation of $\mathbf{R}^{n+1}_i$ is performed at $t^n$, with the inviscid flux Jacobian matrix calculated from a modified numerical inviscid flux in which an extra viscous term \citep{OteroAcc2015,Otero2015,Feng_2021,Wang_2022} is appropriately added to improve the diagonal dominance of the Jacobian matrix. Although incorporating the temporal linearisation of spatial derivatives of solution variables evaluated at the cell centroid into account could be more accurate, it is very complex to do this for these derivatives calculated using the LSQ method in 3D simulations. Therefore, we compromise by treating these spatial derivatives as constant values taken from the temporally closest available point during the linearisation procedure. Consequently, we have
   \begin{equation}\label{ApproximateLinearization}
   \mathbf{R}_i^{n+1}\approx\mathbf{R}_i^{n}+\left(\frac{\partial{\mathbf{R}^{'}_i}}{\partial{\mathbf{U}_i}}\right)^{n}\left(\mathbf{U}_i^{n+1}-\mathbf{U}_i^{n}\right)+\sum\limits_{j=1}^6\left(\frac{\partial{\mathbf{R}^{'}_i}}{\partial{\mathbf{U}_j}}\right)^{n}\left(\mathbf{U}_j^{n+1}-\mathbf{U}_j^{n}\right) \,,
   \end{equation}
   where $\mathbf{R}^{'}_i$ means the residual operator on cell $i$, calculated using the modified numerical inviscid flux \citep{Wang_2022}.
   The resulting linearised system is described as follows:
   \begin{equation}\label{BDF2linearized}
\begin{aligned}
\left(\frac{3 V_i}{2 \Delta t}\mathbf{I}+\left(\frac{\partial\mathbf{R}_i^{'}}{\partial\mathbf{U}_i}\right)^{n}\right)\left(\mathbf{U}_i^{n+1}-\mathbf{U}_i^{n}\right)=- \mathbf{R}^{n}_i+\frac{V_i}{2\Delta t}\left(\mathbf{U}_i^{n}-\mathbf{U}_i^{n-1}\right)
-\left(\frac{\partial\mathbf{R}_i^{'}}{\partial\mathbf{U}_j}\right)^{n}\left(\mathbf{U}_j^{n+1}-\mathbf{U}_j^{n}\right), \quad i\in 1,2,...,N
\end{aligned} \,,
\end{equation}
where $N$ is the number of grid cells in the computational domain. Eq.~(\ref{BDF2linearized}) is then solved using a parallel Lower-Upper Symmetric Gauss-Seidel (LU-SGS) solver \citep{Feng_2021,Wang_2022}, which effectively synchronizes MPI data communication across processors \citep{Feng_2021,WangSubmitted}.

   \subsubsection{PP measure at onset of CME simulation}\label{sec:PPforCME}
   Given that inserting the flux-rope magnetic field into the background corona may cause an extremely high Alfv{\'e}nic speed, and that the filament has a higher density than the background corona, we appropriately increase the plasma density where and when the flux rope is inserted. It is similar to the PP approach in \cite{W_SubmittedCOCONUT} and \cite{wang2025sipifvmtimeevolvingcoronalmodel} and described as: 
  \begin{equation}\label{RhoInner}
  \rho_i=\Upsilon_{\rho_i}\frac{\mathbf{B}_i^2}{V^2_{A,max}}+\left(1-\Upsilon_{\rho_i}\right)\rho_{o,i} \,.
  \end{equation}
  The subscript ``$_o$" refers to the variable in the quasi-steady-state background corona, $\Upsilon_{\rho_i}=0.5+0.5\cdot \tanh\left(\frac{V_{A,i}-V_{A,max}}{V_{fac}}\cdot \pi\right)$ with $V_{fac}=\frac{V_{A,max}}{1000}$, $V_{A,max}=2\frac{\left|\mathbf{B}_{max}\right|}{\rho_{s}^{0.5}}$ with $\left|\mathbf{B}_{max}\right|$ and $\rho_{s}$ being the maximum magnetic field strength in the background corona and the inner boundary plasma density, and $V_{A,i}=\frac{\left|\mathbf{B}_{o,i}+\mathbf{B}_{FR,i}\right|}{\rho_{o,i}^{0.5}}$.
  This does not necessarily represent a physically self-consistent equilibrium, but facilitates a smoother transition from the background corona to the CME configuration and a more stable simulation in the early stage of CME evolution.

  \subsubsection{ESDIRK2 and BDF2 pseudo-time marching method}\label{sec:temporalintegration} 
   For the subsequent time-accurate CME simulation, we employ the ESDIRK2 temporal integration method \citep{KennedyCarpenter2016} to solve Eq.~(\ref{TemporalIntegrationEq}) as follows \citep{BIJL2002313,Lingquan2019}:
   \begin{equation}\label{ESDIRK2Intergration}
   \begin{aligned}
   \mathbf{U}_i^{(1)} &= \mathbf{U}_i^n \\
   \frac{V_i}{\Delta t} \mathbf{U}_i^{(k)} &= \frac{V_i}{\Delta t} \mathbf{U}_i^n - \sum_{l=1}^{k} a_{k,l} \mathbf{R}^{(l)}_i, \quad k = 2, 3 \\
   \mathbf{U}_i^{n+1} &= \mathbf{U}_i^{(3)}
   \end{aligned} \,.
   \end{equation}
   The subscripts ``$^{(k)}$", $k=$ 1, 2, 3 refer to the corresponding values at the $k$-th stage of the ESDIRK2 method, and the coefficients $a_{k,l}$ are listed in Table \ref{tab:esdirk2} \citep{KennedyCarpenter2016}, where $b_2=\frac{1-2\kappa}{4\kappa}$ and $\kappa=1-\frac{\sqrt{2}}{2}$. 
\begin{table}[htbp]
\centering
\[
\begin{array}{c|ccc}
a_{1,l}       & 0     & 0       & 0       \\
a_{2,l} & \kappa  & \kappa  & 0       \\
a_{3,l}       & 1 - b_2 - \kappa  & b_2  & \kappa  \\
\hline
b_i     & 1 - b_2 - \kappa   & b_2     & \kappa 
\end{array}
\]
\caption{Butcher table of a 2nd-order, three-stage ESDIRK2 method.}
\label{tab:esdirk2}
\end{table}

To improve temporal accuracy in CME simulations, a pseudo time $\tau$ is introduced at the 2nd and 3rd stages of the ESDIRK2 method, allowing the solution at these stages to be obtained by solving a steady-state problem on $\tau$. Consequently, we get the following equations:
\begin{equation}\label{ESDIRK2IntergrationPsudoTime}
V_{i}\frac{\partial \mathbf{U}_{i}^{(k)}}{\partial \tau}+\left(\frac{V_i}{\Delta t} \mathbf{U}_i^{(k)} - \frac{V_i}{\Delta t} \mathbf{U}_i^n + \sum_{l=1}^{k} a_{k,l} \mathbf{R}^{(l)}_i\right)= \mathbf{0} , \quad k = 2, 3 \,.
\end{equation}
By applying the BDF2 scheme \citep{Nived2023} to solve Eq.~(\ref{ESDIRK2IntergrationPsudoTime}) and performing an approximate temporal linearisation, similar to Eq.~(\ref{ApproximateLinearization}), for $\mathbf{R}^{(l)}_i$, we get the following equations:
\begin{equation}\label{BDF2Psudori}
\begin{aligned}
&V_i\frac{3\mathbf{U}_i^{(k,m+1)}-4\mathbf{U}_i^{(k,m)}+\mathbf{U}_i^{(k,m-1)}}{2~\Delta \tau^k}+V_i\frac{\mathbf{U}_i^{(k,m+1)}-\mathbf{U}_i^{(k,m)}}{\Delta t}
+V_i\frac{\mathbf{U}_i^{(k,m)}-\mathbf{U}_i^n}{\Delta t}+a_{k,k}\left(\frac{\partial\mathbf{R}_i^{'}}{\partial\mathbf{U}_i}\right)^{(k,m)}\left(\mathbf{U}_i^{(k,m+1)}-\mathbf{U}_i^{(k,m)}\right)=\\ 
&-a_{k,k}\left(\frac{\partial\mathbf{R}_i^{'}}{\partial\mathbf{U}_j}\right)^{(k,m)}\left(\mathbf{U}_j^{(k,m+1)}-\mathbf{U}_j^{(k,m)}\right)
-a_{k,k} \mathbf{R}^{(k,m)}_i
-\sum_{l=1}^{k-1} a_{k,l} \mathbf{R}^{(l)}_i, \quad k = 2, 3 
\end{aligned} \,.
\end{equation}
The subscript ``$^{(k,m)}$" refers to the corresponding variables at the $m$-th pseudo-time level during the $k$-th stage of the ESDIRK2 method, and $\Delta \tau^k= \tau^{(k,m+1)}- \tau^{(k,m)}$. 

Obviously, as the value of $a_{k,k}$
approaches zero, Eq.~(\ref{BDF2Psudori}) becomes an explicit time integration scheme, which becomes unstable for large time steps. 
Furthermore, inspired by previous research work in which the numerical stability of implicit solvers was improved by artificially amplifying the diagonal block of the Jacobian matrix \citep{ASADA2018249,Liu_2023,Yasue2010}, and considering that our purpose is only to reach a steady state in $\tau$, we increase the coefficient of the diagonal block component of the implicit matrix
$\left(\frac{\partial\mathbf{R}_i^{'}}{\partial\mathbf{U}}\right)^{(k,m)}$ from $a_{k,k}$ to 1.1, as described below.
\begin{equation}\label{BDF2Psud}
\begin{aligned}
&V_i\frac{3\mathbf{U}_i^{(k,m+1)}-4\mathbf{U}_i^{(k,m)}+\mathbf{U}_i^{(k,m-1)}}{2~\Delta \tau^k}+V_i\frac{\mathbf{U}_i^{(k,m+1)}-\mathbf{U}_i^{(k,m)}}{\Delta t}
+V_i\frac{\mathbf{U}_i^{(k,m)}-\mathbf{U}_i^n}{\Delta t}+1.1~\left(\frac{\partial\mathbf{R}_i^{'}}{\partial\mathbf{U}_i}\right)^{(k,m)}\left(\mathbf{U}_i^{(k,m+1)}-\mathbf{U}_i^{(k,m)}\right)=\\ 
&-a_{k,k}\left(\frac{\partial\mathbf{R}_i^{'}}{\partial\mathbf{U}_j}\right)^{(k,m)}\left(\mathbf{U}_j^{(k,m+1)}-\mathbf{U}_j^{(k,m)}\right)
-a_{k,k} \mathbf{R}^{(k,m)}_i
-\sum_{l=1}^{k-1} a_{k,l} \mathbf{R}^{(l)}_i, \quad k = 2, 3 
\end{aligned} \,.
\end{equation}

From Eq.~(\ref{BDF2Psud}), we have 
\begin{equation}\label{BDF2Psudlinearized}
\begin{aligned}
&\left(\frac{3 V_i}{2 \Delta \tau^k}\mathbf{I}+\frac{V_i}{\Delta t}\mathbf{I}+1.1~\left(\frac{\partial\mathbf{R}_i^{'}}{\partial\mathbf{U}_i}\right)^{(k,m)}\right)\left(\mathbf{U}_i^{(k,m+1)}-\mathbf{U}_i^{(k,m)}\right)=
\frac{V_i}{2\Delta \tau^k}\left(\mathbf{U}_i^{(k,m)}-\mathbf{U}_i^{(k,m-1)}\right)+\frac{V_i}{\Delta t}\left(\mathbf{U}_i^{n}-\mathbf{U}_i^{(k,m)}\right)-\sum_{l=1}^{k-1} a_{k,l} \mathbf{R}^{(l)}_i-\\
&a_{k,k} \mathbf{R}^{(k,m)}_i-a_{k,k}\left(\frac{\partial\mathbf{R}_i^{'}}{\partial\mathbf{U}_j}\right)^{(k,m)}\left(\mathbf{U}_j^{(k,m+1)}-\mathbf{U}_j^{(k,m)}\right), \quad k = 2, 3 
\end{aligned} \,.
\end{equation}
Consequently, from Eq.~(\ref{BDF2Psudlinearized}) we get the following linearised system:
\begin{equation}\label{BDF2PseIBElinearized}
\mathbf{A}_{8N\times8N}\left(\Delta \mathbf{U}^{(k,m)}\right)_{8N}=\mathbf{b}_{8N}, \quad k = 2, 3 \,,
\end{equation}
   where $\mathbf{A}_{8N\times8N}=\left(\frac{3V}{2\Delta \tau}\mathbf{I}+\frac{V}{\Delta t}\mathbf{I}+a_{k,k}\left(\frac{\partial\mathbf{R}^{'}}{\partial\mathbf{U}}\right)_{L\big/U}^{k,m}+1.1~\left(\frac{\partial\mathbf{R}^{'}}{\partial\mathbf{U}}\right)_{D}^{k,m}\right)_{8N\times8N}$, $\left(\Delta \mathbf{U}^{(k,m)}\right)_{8N}=\left(\mathbf{U}^{(k,m+1)}-\mathbf{U}^{(k,m)}\right)_{8N}$ and $\mathbf{b}_{8N}=\frac{V}{2\Delta \tau^k}\left(\mathbf{U}^{(k,m)}-\mathbf{U}^{(k,m-1)}\right)_{8N}+\frac{V}{\Delta t}\left(\mathbf{U}^{n}-\mathbf{U}^{(k,m)}\right)_{8N}-\sum_{l=1}^{k-1} a_{k,l} \left(\mathbf{R}^{(l)}\right)_{8N}-a_{k,k} \left(\mathbf{R}^{(k,m)}\right)_{8N}$ with $N$ being the number of grid cells in the computational domain. Meanwhile, $\left(\frac{\partial\mathbf{R}^{'}}{\partial\mathbf{U}}\right)_{L\big/U}^{k,m}$ denotes the lower and upper triangle block components of $\left(\frac{\partial\mathbf{R}^{'}}{\partial\mathbf{U}}\right)^{k,m}$, and $\left(\frac{\partial\mathbf{R}^{'}}{\partial\mathbf{U}}\right)_{D}^{k,m}$ denotes the diagonal block component of $\left(\frac{\partial\mathbf{R}^{'}}{\partial\mathbf{U}}\right)^{k,m}$.
   
   It is noticed that the solution prior to the current pseudo-time level is also required in the BDF2 pseudo-time marching method; however, $\mathbf{U}_i^{(k,-1)}$ is missing for the initial pseudo-time level. To address this, we set $\mathbf{U}_i^{(k,0)}=\mathbf{U}_i^{(k-1)}$ and employ the backward Euler method to calculate $\mathbf{U}_i^{(k,1)}$ as described in Eq.~(\ref{BackwardEuler}), and then employ BDF2 as described in Eq.~(\ref{BDF2Psud}) to update the solution in the following pseudo-time levels.  
\begin{equation}\label{BackwardEuler}
\begin{aligned}
&\left(\frac{V_i}{\Delta \tau^k}\mathbf{I}+\frac{V_i}{\Delta t}\mathbf{I}+1.1~\left(\frac{\partial\mathbf{R}_i^{'}}{\partial\mathbf{U}_i}\right)^{(k,0)}\right)\left(\mathbf{U}_i^{(k,1)}-\mathbf{U}_i^{(k,0)}\right)= 
V_i\frac{\mathbf{U}_i^n-\mathbf{U}_i^{(k,0)}}{\Delta t}-a_{k,k}\left(\frac{\partial\mathbf{R}_i^{'}}{\partial\mathbf{U}_j}\right)^{(k,0)}\left(\mathbf{U}_j^{(k,1)}-\mathbf{U}_j^{(k,0)}\right)-\\
&-\sum_{l=1}^{k-1} a_{k,l} \mathbf{R}^{(l)}_i-a_{k,k} \mathbf{R}^{(k,0)}_i, \quad k = 2, 3; \quad i\in 1,2,...,N
\end{aligned} \,.
\end{equation}
   
   We solve Eq.~(\ref{BackwardEuler}) and Eq.~(\ref{BDF2PseIBElinearized}) by the parallel LU-SGS solver \citep{Feng_2021,Wang_2022}. Once the simulation reaches a steady-state condition \citep{WangSubmitted} at $\tau^k$, we set $\left(\mathbf{U}^{(k)}\right)_{8N}=\left(\mathbf{U}^{(k,m+1)}\right)_{8N}$ and stop the pseudo-time marching at the $k$-th stage.

 \subsubsection{Time steps in background and CME simulations}\label{sec:TimeStepAdjustment}
  Since the time-step size $\Delta t$ can significantly influence solution accuracy, computational efficiency, and numerical stability in coronal and CME simulations \citep{WangSubmitted, W_SubmittedCOCONUT, wang2025COCONUTMayEvent}, it is essential to carefully select an appropriate $\Delta t$ that balances all three aspects.
 
  During the quasi-steady-state coronal simulation, the magnetic and flow fields slowly evolve toward equilibrium. Increasing $\Delta t$ typically enhances computational efficiency without compromising the final solution accuracy. As in \cite{Wang_2022}, we gradually increase the time step $\Delta t$ from a time step constrained by the CFL stability condition \citep{FengandLiu2019,Xia_2018}, denoted by $\Delta t^{\rm expl}$, to a reference time scale $\tau_{flow}$, as defined in \cite{Feng_2021}. In this paper, $\tau_{flow}$ is around 3.14 minutes, which is about 470 times larger than the time step constrained by the CFL stability constraint, demonstrating a significant advantage of the implicit method over explicit schemes. We now present the formulation of $\Delta t^{\rm expl}$.
   \begin{equation}\label{explicittimestep}
   \begin{aligned}
\Delta t^{\rm expl}&=0.5\min\left(\Delta t_d,\Delta t_c\right)\\
\Delta t_c&=\min\limits_{\forall {\rm cell}_{i}} \frac{\Delta h_i} {\max\limits_{\forall \rm faces} \left(\left|v_n\right|+c_{f}\right)}\\ \Delta t_d&=\min\limits_{\forall {\rm cell}_{i}} \frac{\Delta h_i^2} {\left(\gamma-1\right)\xi T^{2.5}\max\limits_{\forall \rm faces} c_{An}^2}
\end{aligned}
\end{equation}
Here $\Delta h_i$ is the diameter of the smallest inscribed sphere in ${\rm cell}_i$, $v_n$, $c_{f}$ and $c_{An}$ are the normal plasma velocity, normal fast magnetosonic speed and normal Alfv{\'e}n speed on ${\rm cell}_i$'s surface, respectively. $T$ is the temperature of the bulk plasma and $\xi T^{2.5}$ denotes the thermal conductivity coefficient defined the same as in \cite{wang2025sipifvmtimeevolvingcoronalmodel}.

  As for CME simulations, the magnetic and flow fields evolve much more rapidly, and because the temporal evolution of CMEs is of particular interest, it is essential to ensure required temporal accuracy. Therefore, we employ a relatively small $\Delta t$ in addition to using the ESDIRK2 temporal integration scheme combined with the BDF2 pseudo-time marching method. Moreover, manually inserting a flux rope into the background corona often leads to significant imbalance and high-frequency transients at the onset of CME simulations, a large $\Delta t$ may fail to resolve the high-frequent evolution and result in numerical instability. 
  Therefore, we set $\Delta t$ as the value constrained by the CFL stability limitation for explicit schemes during the first 50 physical time steps of the CME simulation, and then gradually increase $\Delta t$ to 0.125 $\tau_{flow}$, which is about 0.4 minutes.

  Moreover, the size of the pseudo-time step $\Delta \tau^k$ affects the convergence rate of steady-state simulations in the pseudo time during the $k$-th stage of the ESDIRK2 method. A small $\Delta \tau^k$ leads to slow convergence, while an extremely large $\Delta \tau^k$ may cause numerical instability. Moreover, an infinitely large $\Delta \tau^k$ is equivalent to Newton's method \citep{Lingquan2019}, which offers faster convergence but is sensitive to the initial guess of the solution. 
  Consequently, we set $\Delta \tau^k = 10^{20}$ for the first pseudo-time step and $\Delta \tau^k = 10 \Delta t$ for the following pseudo-time steps to ensure both accurate and rapid convergence to the steady-state solution in pseudo time. Meanwhile, we adopt the same criteria as in \cite{WangSubmitted} to determine whether the simulation has reached a steady state, and we limit the number of pseudo-time steps in each stage of ESDIRK2 to a maximum of five to further improve computational efficiency.

\section{Numerical results}\label{sec:Numerical Results}
   In this paper, we first perform a quasi-steady-state coronal simulation using the BDF2 temporal integration scheme, as described in Eq.~(\ref{BDF2linearized}), to obtain the background coronal structure on 21 Jun 2011, the day of the CME event. This data falls within CR 2111, which spans from June 5, 2011 to July 3, 2011 and is near the solar maximum of Solar Cycle 24. We then manually insert an observation-based RBSL flux rope to initiate a CME event and simulate its evolution and propagation from the solar surface to 0.1 AU using the ESDIRK2 method, coupled with the BDF2 pseudo-time marching approach described in  Eq.~(\ref{BDF2PseIBElinearized}).
   During the simulations, the maximum magnetic field strength near the solar surface is approximately 51 Gauss before and 58 Gauss after inserting the flux rope. The corresponding minimum plasma values of $\beta$ are about $5\times10^{-4}$ and $8\times10^{-4}$, respectively. 

   In this work, all simulations were performed using 192 CPUs in the Genius cluster of the Tier-2 supercomputer at the Vlaams Supercomputer Centrum\footnote{\url{https://www.vscentrum.be/}}. The quasi-steady-state coronal simulation converged in 1342 iterations, taking 20 minutes of computational time. The subsequent simulation of 6.1 hours of CME evolution was completed in 925 physical time steps, taking 1.46 hours of computational time. 

   \subsection{Comparison of the conventional and the extended magnetic field decomposition method}\label{sec:ComparisonQuasisteadystatesimulation}
   To further validate the extended magnetic field decomposition method, we also performed simulations using the conventional magnetic field decomposition approach where the magnetic field is decomposed into a static magnetic field $\mathbf{B}_0$ and a time-varying field $\mathbf{B}_1$ \citep{Tanaka1995,Powell1999,FUCHS2010JCP,Guo2015,Xia_2018,Feng_2021,Wang_2022}. The only difference from the algorithm described in Section~\ref{sec:Description of the novel MHD Model} is that $\mathbf{B}_0$ remains a static potential field throughout the simulation. Although the conventional approach also performs well in quasi-steady-state coronal simulations, it fails at the $7$-th physical time step of the CME simulation due to the fact that $\mathbf{B}_1$ is no longer small after inserting the flux rope into the background corona. 
   As for the quasi-steady-state background coronal simulation, we stop at approximately the same relaxation time as that calculated by the extended magnetic field method. It required 1343 iterations and also took 20 minutes of computational time. In the following context of this subsection, we compare the simulation results calculated using the conventional and extended magnetic field decomposition approaches, respectively.

\begin{figure*}[htpb!]
\begin{center}
  \vspace*{0.01\textwidth}
    \includegraphics[width=0.9\linewidth,trim=1 1 1 1, clip]{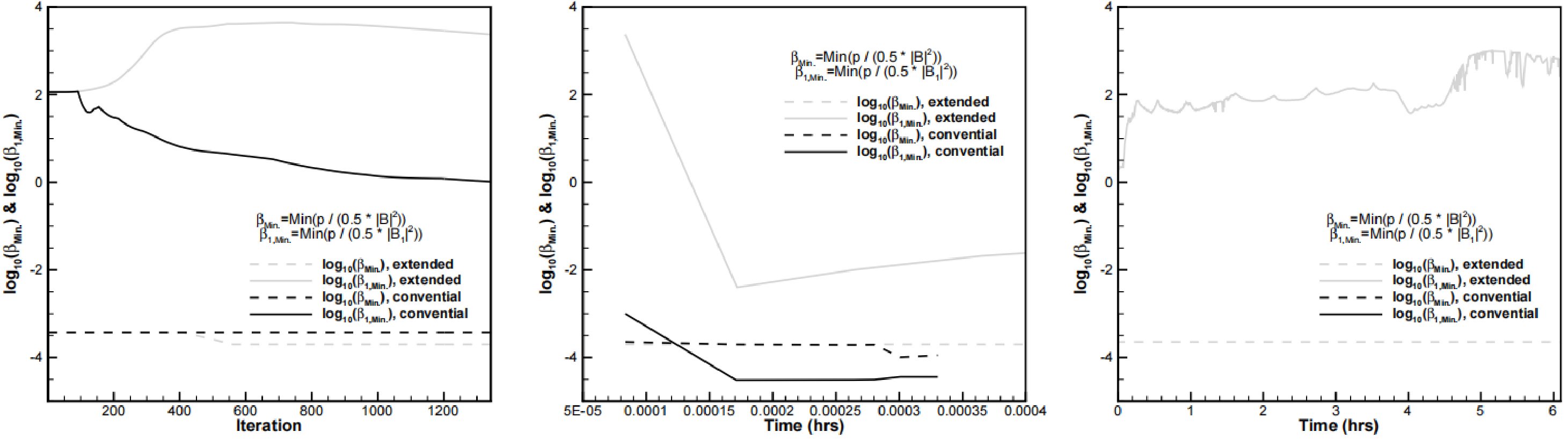}
\end{center}
\caption{
Evolution of the base-10 logarithms of the minimum values of $\beta_1 = \frac{p}{0.5\,|\mathbf{B}_1|^2}$ and $\beta = \frac{p}{0.5\,|\mathbf{B}|^2}$ within the entire computational domain, denoted by $\log_{10}(\beta_{1,\mathrm{min}})$ and $\log_{10}(\beta_{\mathrm{min}})$, respectively. The left and right panels denote the evolution during the quasi-steady-state coronal simulation and the time-varying CME simulation, along with a close-up view of the profile during the first $4 \times 10^{-4}$ hours of the CME simulation (middle). The solid and dashed lines represent the evolution of $\log_{10}(\beta_{1,\mathrm{min}})$ and $\log_{10}(\beta_{\mathrm{min}})$, respectively, while gray and black lines indicate results calculated using the extended and conventional magnetic field decomposition approaches.
}\label{MinBetaEvolution}
\end{figure*}
   Since the thermal pressure is derived by $p=\left(\gamma-1\right)\cdot\left(E_1-\frac{1}{2}\rho\mathbf{v}^2-\frac{1}{2}\mathbf{B}_1^2\right)$, it is important to ensure that the ratio $\beta_1=\frac{p}{0.5\cdot\mathbf{B}_1^2}$ is not too small. Otherwise, catastrophic cancellation may occur, where significant digits are lost due to round-off errors in floating-point calculations. This can result in nonphysical negative pressure values and ultimately cause the MHD code to crash.  
   In Fig.~\ref{MinBetaEvolution}, we show the evolution of $\log_{10}(\beta_{1,\mathrm{min}})$ and $\log_{10}(\beta_{\mathrm{min}})$ calculated by the conventional and extended magnetic field decomposition approaches, denoted by the solid and dashed black and gray lines, respectively. $\beta_{1,\mathrm{min}}=\min(\beta_1)$ and $\beta_{\mathrm{min}}=\min(\beta)$ mean the minimum values of $\beta_1=\frac{p}{0.5\cdot\mathbf{B}_1^2}$ and $\beta=\frac{p}{0.5\cdot\mathbf{B}^2}$ within the computational domain. The left and right panels correspond to the quasi-steady-state coronal simulation and the time-dependent CME simulation, respectively, and the midle panel shows a close-up view of the profile during the first $4 \times 10^{-4}$ hours of the CME simulation. 
   
   It can be seen that $\beta_{\rm min}$ remains around $10^{-3.5}$ in both the quasi-steady-state coronal simulation and the time-dependent CME simulation. In the quasi-steady-state case, $\beta_{1,\min}$ gradually decreases to about 1 when using the conventional magnetic field decomposition method, whereas it remains above $10^2$ and eventually increases to approximately $10^4$ with the extended decomposition method. In the CME simulation, $\beta_{1,\min}$ drops below $10^{-4}$ and the simulation quickly crashes when using the conventional method. In contrast, although $\beta_{1,\min}$ initially decreases to about $10^{-2.5}$ with the extended method, it rapidly rises to around $10^2$ in subsequent time steps, thereby maintaining numerical stability.

\begin{figure*}[htpb!]
\begin{center}
  \vspace*{0.01\textwidth}
    \includegraphics[width=0.9\linewidth,trim=1 1 1 1, clip]{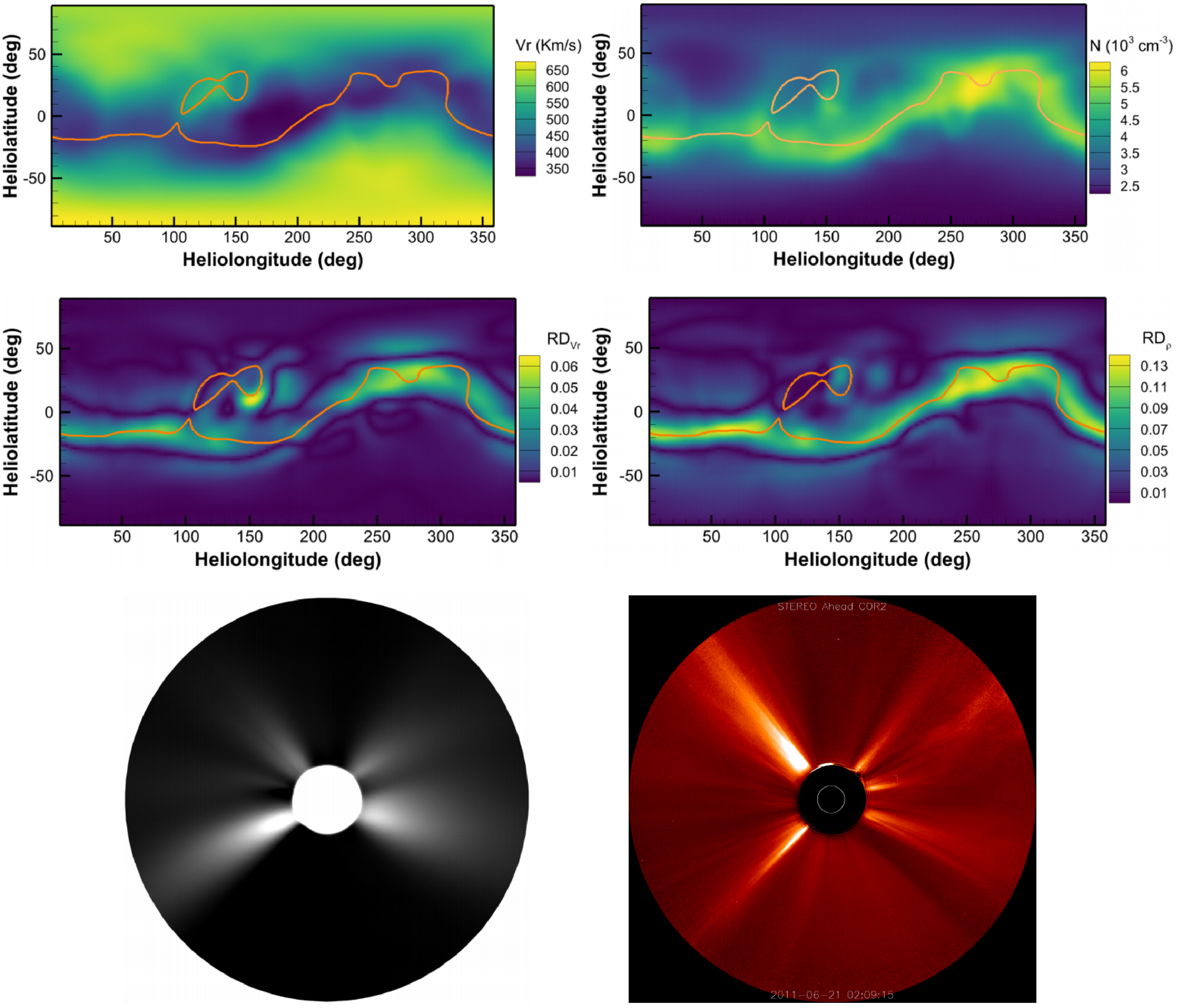}
\end{center}
\caption{
Distribution of the radial plasma velocity $V_r$, in unit of $\rm km~ s^{-1}$, and proton number density $N$, in unite of $10^3~\rm cm^{-3}$, at 20 $R_s$ (top). The middle panels show the relative differences in radial velocity and plasma density, denoted by $\rm RD_{Vr}$ and $\rm RD_{\rho}$ , respectively, between the coronal simulation results calculated by the conventional and extended magnetic field decomposition approaches. The orange lines mean MNLs ($B_r=0$). The bottom panels present the white-light pB image ranging from $2.5~R_s$ to $15~R_s$ synthesized from the simulation results calculated by the extended magnetic field decomposition method (bottom left), and the corresponding observed pB image from COR2/STEREO-A (bottom right).
}\label{CoronalMeridian}
\end{figure*}
  In Fig.~\ref{CoronalMeridian}, we compare the radial velocity and plasma density calculated by the extended and conventional magnetic field decomposition methods (top and middle), as well as the simulated and observed white-light polarized brightness (pB) images (bottom) of the background corona. The top row shows the simulated radial velocity $V_r \left( {\rm km~ s^{-1}}\right)$ and plasma density $N \left(10^3~\rm cm^{-3}\right)$ calculated using the extended magnetic field decomposition method. The middle row displays the relative differences in radial velocity (left), ${\rm RD}_{V_r}$, and plasma density (right), ${\rm RD}_{\rho}$, between the background coronal results calculated by the conventional and extended magnetic field decomposition approaches. Here
  $${\rm RD}_{V_r} = \frac{\left|V_r^{\rm EXT} - V_r^{\rm CONV}\right|}{\left|V_r^{\rm EXT}\right|}\,,$$
  $${\rm RD}_{\rho} = \frac{\left|\rho^{\rm EXT} - \rho^{\rm CONV}\right|}{\rho^{\rm EXT}}\,.$$  The subscripts ``$^{\rm EXT}$" and ``$^{\rm CONV}$" denote the corresponding variables calculated by the extended and conventional magnetic field decomposition methods, respectively. The orange lines overlaid on these contours denote the magnetic neutral lines (MNLs). The bottom row presents the pB images ranging from 2.5 $R_s$ to 15 $R_s$ synthesized from the simulation results (bottom left) calculated by the extended magnetic field decomposition method and observed (bottom right) by the innermost coronagraph of the Sun Earth Connection Coronal and Heliospheric Investigation (SECCHI) instrument suite onboard the \textit{Solar Terrestrial Relations Observatory Ahead} (STEREO-A) spacecraft\footnote{\url{https://stereo-ssc.nascom.nasa.gov/browse/}} \citep{Howard2008}. 

  It can be seen that the simulated high-density, low-speed flows are primarily concentrated around the MNLs. The relative differences in radial velocity ${\rm RD}_{V_r}$ and plasma density ${\rm RD}_{\rho}$ at 20 $R_s$ between the results calculated by conventional and extended magnetic field decomposition methods are generally below $2\%$ and $5\%$, respectively, with larger discrepancies mainly distributed around the MNLs. It shows that the simulation results calculated by both methods are consistent with each other. Furthermore, the simulated pB image successfully reproduces the prominent bright structures centered around $40^{\circ}$ N and $15^{\circ}$ S on the west limb, as well as the structure centered around $46^{\circ}$ N on the east limb. However, the observed bright structure centered around $40^{\circ}$ S on the east limb appears shifted northward by about $15^{\circ}$ in the simulation, and a small bright feature along the equator on the west limb seen in observations is absent from the simulation. The discrepancy may be attributed to the lack of accurate magnetic field measurements in the polar regions, the relatively coarse mesh resolution fails to capture fine-scale magnetic field structures available in observation, and the use of the synoptic magnetogram, where magnetic field data of different longitudes are observed at different times \citep{Hamada2018}, fails to reflect the true magnetic field distribution at the corresponding physical time of the global coronal simulation.

\begin{table}
\caption{Average relative differences between background coronal simulations calculated by the extended and conventional magnetic field decomposition approaches.}
\label{diffaveragy}
\resizebox{0.8\linewidth}{!}{
\begin{tabular}{llll}
\hline\noalign{\smallskip}
 Parameters  &   ${\rm RD}_{{\rm ave},\rho}$  & ${\rm RD}_{{\rm ave},{V_r}}$ & ${\rm RD}_{{\rm ave},{\left|\mathbf{B}\right|}}$\\
\noalign{\smallskip}\hline\noalign{\smallskip}
  magnitude & 2.21\% & 1.06\% & 0.92\%\\
\noalign{\smallskip}\hline
\end{tabular}
}
\end{table}
  In Table \ref{diffaveragy}, we further list the average relative differences in proton number density, ${\rm RD}_{{\rm ave},\rho}$, radial velocity, ${\rm RD}_{{\rm ave},{V_r}}$, and magnetic field strength, ${\rm RD}_{{\rm ave},{\left|\mathbf{B}\right|}}$, between the background coronal simulation results calculated by the extended and conventional magnetic field approaches, respectively. Here, 
  $${\rm RD}_{{\rm ave},\rho}=\sum\limits_{i=1}^N\big|\rho_i^{\rm EXT}-\rho_i^{\rm CONV}\big|\big/\sum\limits_{i=1}^N\rho_i^{\rm EXT} \,,$$
  $${\rm RD}_{{\rm ave},{V_r}}=\sum\limits_{i=1}^N\big|{V_r}_i^{\rm EXT}-{V_r}_i^{\rm CONV}\big|\big/\sum\limits_{i=1}^N {V_r}_i^{\rm CONV}\,,$$  
  $${\rm RD}_{{\rm ave},\left|\mathbf{B}\right|}=\sum\limits_{i=1}^N\big|\mathbf{B}_i^{\rm EXT}-\mathbf{B}_i^{\rm CONV}\big|\big/\sum\limits_{i=1}^N\left|\mathbf{B}\right|_i^{\rm EXT}\,.$$ 
  The subscripts `$_{i}$' denote the corresponding cell-averaged variable in cell $i$, and $N$ is the number of cells in the computational domain. It can be seen that the average relative differences in proton number density, radial velocity, and magnetic field strength between the coronal simulation results obtained using the extended and conventional magnetic field decomposition methods are only $2.21\%$, $1.06\%$, and $0.92\%$, respectively, further validating the accuracy of the extended magnetic field decomposition method. 

  \subsection{Evolution of magnetic field structure}\label{sec:Quasisteadystatesimulation}
   In the sub-Alfv{\'e}nic coronal region, the magnetic field dominates the dynamics of coronal structures.  This subsection presents the evolution of open-field regions and the radial magnetic field near the solar surface. The evolution of some selected magnetic field lines surrounding the flux rope is also illustrated.

  \subsubsection{Evolution of open-field regions and radial magnetic field near solar surface}\label{sec:Distributions of the open-field regions}
\begin{figure*}[htpb!]
\begin{center}
  \vspace*{0.01\textwidth}
\includegraphics[width=0.9\linewidth,trim=1 1 1 1, clip]{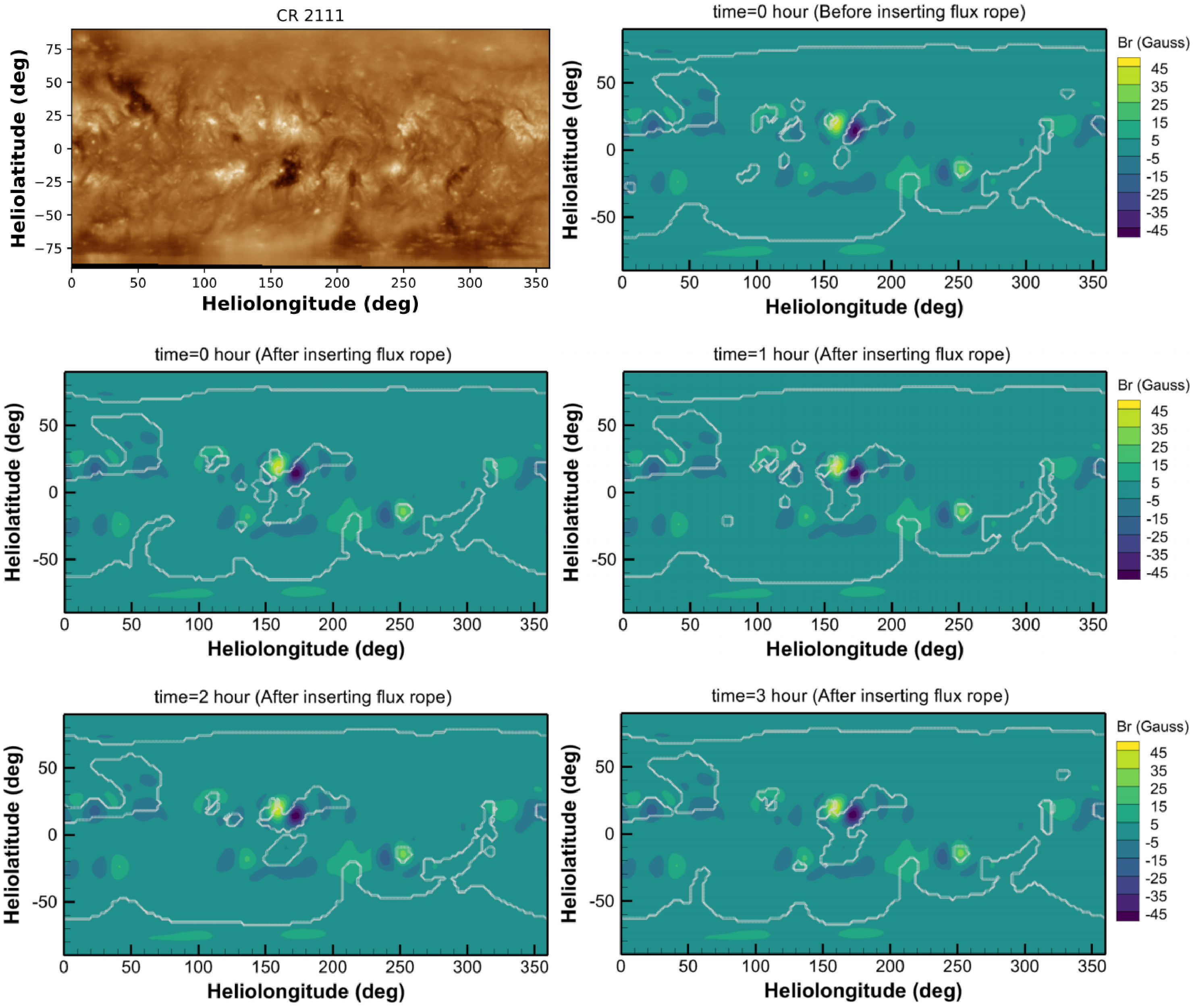}
\end{center}
\caption{Distribution of the radial magnetic field at 1.015 $R_s$ in the background corona (top right), and at 0, 1, 2, and 3 hours (middle and bottom) of the CME simulation, alongside the synoptic map of extreme ultraviolet (EUV) observations from the 193~\AA\ channel of AIA onboard SDO for CR 2111 (top left). The white lines overlaid on the radial magnetic field contours indicate the boundaries of the open-field regions.}\label{magnetCHsurface}
\end{figure*}
  Magnetic field lines originating from coronal holes (CHs) are typically open to interplanetary space, resulting in regions of low plasma density. Consequently, CHs appear as dark areas in images captured in extreme ultraviolet (EUV) and soft X-ray wavelengths. Given that the distribution of CHs varies across different phases of the solar activity cycle and are widely recognized as the primary source region of high-speed solar wind flows \citep{Nolte1976,CRANMER20023,Hofmeister2022}, they are considered one of the most prominent and dynamic features of the solar corona \citep{FengMa2015,Feng_2017,Frazin2007,Hayes2001,Linker1999JGR,Petrie2011SoPh}. In Fig.~\ref{magnetCHsurface}, we illustrate evolution of the radial magnetic field near the solar surface, actually at $1.015~R_s$, and the corresponding open-field regions, as well as the synoptic map of the EUV observations from the 193~\AA\ channel of the Atmospheric Imaging Assembly (AIA) telescope on board the Solar Dynamics Observatory \citep{Lemen2012}\footnote{\url{https://sdo.gsfc.nasa.gov/data/synoptic/}} for CR 2111. These images are presented in a co-rotating heliographic coordinate (HGC) system. 
  The synoptic EUV map is generated by concatenating a series of meridional strips extracted from full-disk images over the duration of a complete CR period \citep{Hamada2018}. At the time corresponding to the simulated CME event, the Earth is located approximately $157^{\circ}$ in longitude of the synoptic EUV map. Additionally, we trace magnetic field lines to identify open-field regions near the solar surface and present their boundaries with white lines overlaid on the corresponding radial magnetic field contours.

  It can be seen that the open-field regions from both background corona and the time-evolving CME simulation results capture the observed polar CH $70^{\circ}$ southern poleward. The leading CHs around $220^{\circ}$ and $280^{\circ}$ in longitude, extending from the southern pole to near the solar equator, as well as the isolated CH centered at $(\theta,\phi)=(40^{\circ} {\rm N}, 60^{\circ})$ and detached from the northern polar CH, are also well captured. Furthermore, we focus on the isolated CH centered at $(\theta,\phi) = (0^{\circ}, 150^{\circ})$, near which an active region emerges. This active region appears as a magnetic dipole around $(\theta,\phi) = (20^{\circ}~\mathrm{N}, 160^{\circ})$ in the global magnetogram, into which the observation-based flux rope is inserted. Notably, the open-field region associated with the isolated CH is absent in the background coronal simulation but reappears after the insertion of the flux rope. This highlights the importance of incorporating high-resolution magnetic-field observations in global coronal models to better reproduce realistic coronal structures.
  During the subsequent CME simulation, the scattered low-latitude CHs near the isolated CH undergo rapid evolution, suggesting frequent magnetic reconnection between the CME flux rope field and the ambient coronal magnetic field. These findings indicate the necessity of adopting a more realistic time-evolving background corona, driven by the time-varying inner-boundary magnetic field, to improve the consistency of CME simulations with observations. 

  \subsubsection{Evolution of selected magnetic field lines}\label{sec:Simulated results near the Sun}
\begin{figure*}[htpb!]
\begin{center}     \includegraphics[width=0.8\linewidth,trim=1 1 1 1, clip]{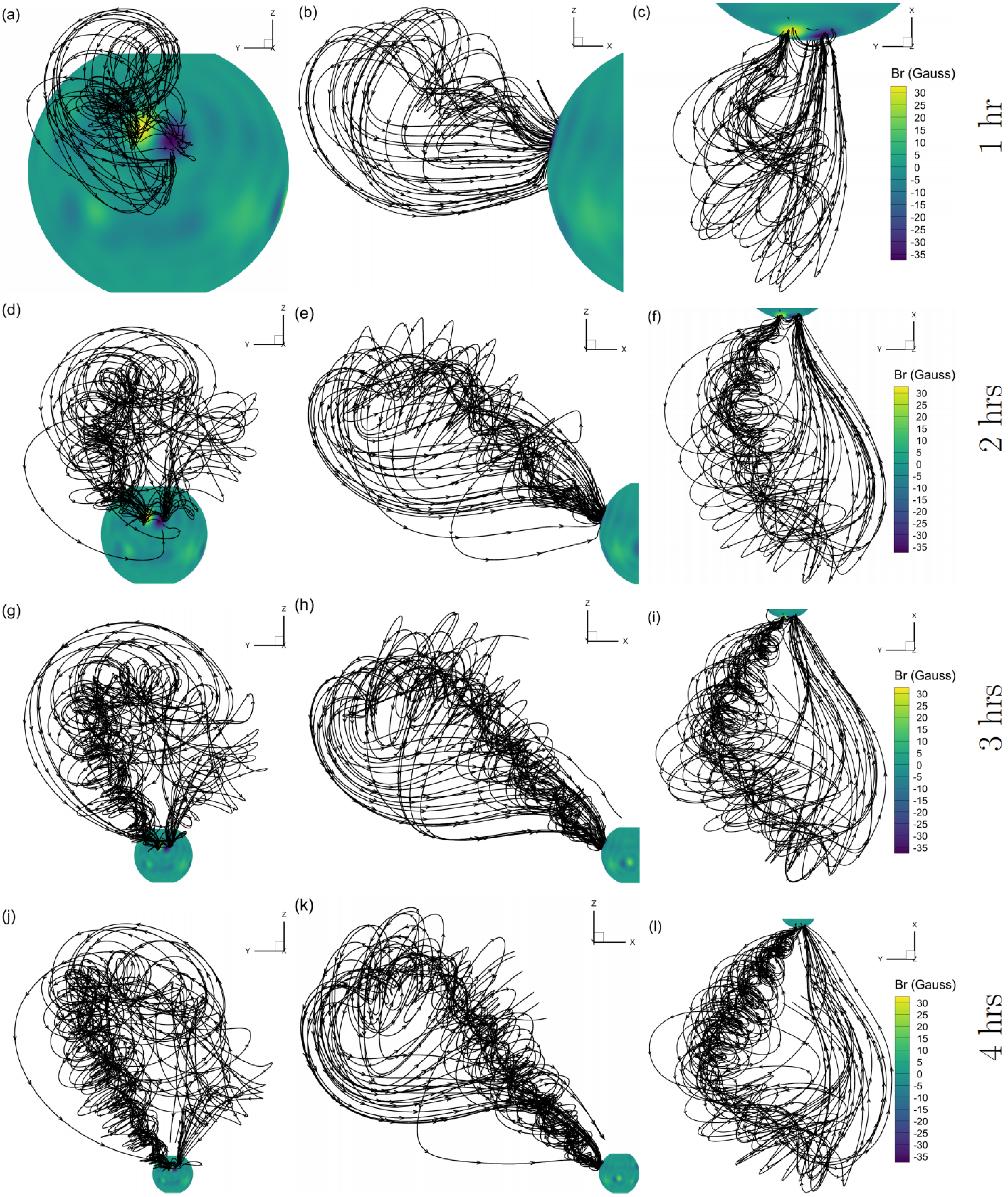}
\end{center}
  \caption{3D view of the magnetic field topology during the CME simulation. The solid lines denote magnetic field lines traced from magnetic field in the region defined by $r \in \left(1.035~R_s, ~20R_s\right)$, $\theta \in \left(22.5^{\circ},~157.5^{\circ}\right)$, and $\phi \in \left(\phi_{FR}-135^{\circ},\phi_{FR}+135^{\circ}\right)$, which encloses the observation-based flux rope. Here $\phi_{FR}=163.8^{\circ}$ denotes the longitude of the midpoint between the two footpoints of the flux rope axis path.
  Rows 1–4 correspond to the CME simulation results at 1, 2, 3, and 4 hours, respectively. The magnetic field lines shown in the left, middle, and right panels are viewed from the directions $\left(\theta,\phi\right)=\left(90^{\circ},\phi_{FR}\right)$, $\left(\theta,\phi\right)=\left(90^{\circ},\phi_{FR}+90^{\circ}\right)$, and along the Z-axis, respectively. These three viewing directions are mutually orthogonal, and the magnetic field lines in each panel are traced from the same set of seed points.}\label{3DMFLs}
\end{figure*}
   In Fig.~\ref{3DMFLs}, we present snapshots of the magnetic field lines at 1, 2, 3, and 4 hours to demonstrate the propagation of the observation-based flux rope in the background coronal structure of CR 2111. These magnetic field lines are traced from the CME simulation results in a region of $r \in \left(1.035~R_s, ~20R_s\right)$, $\theta \in \left(22.5^{\circ},~157.5^{\circ}\right)$, and $\phi \in \left(\phi_{FR}-135^{\circ},\phi_{FR}+135^{\circ}\right)$, where $\phi_{FR}=163.8^{\circ}$ denotes the longitude of the midpoint between the two footpoints of the flux rope axis path. It encloses the flux rope and effectively captures the significant changes in its overall morphology as the CME propagates outward. The magnetic field lines in the left, middle, and right panels are viewed in three orthogonal directions, along $\left(\theta,\phi\right)=\left(90^{\circ},\phi_{FR}\right)$, $\left(\theta,\phi\right)=\left(90^{\circ},\phi_{FR}+90^{\circ}\right)$, and the Z-axis, respectively. Additionally, the magnetic field lines in each panel are traced from the same set of seed points to ensure a consistent representation of their continuous evolution.

   It can be seen that as the volume occupied by the CME flux rope expands outward, its propagation direction gradually shifts to the northeast of the Sun–Earth line. Combined with the initial magnetic field lines shown in Fig.~\ref{Fluxropepath}, which are nearly directed toward Earth, it can be inferred that the CME undergoes a significant northeastward deflection during the first hour of the CME simulation. 
   This result is consistent with observations and the earlier research work by \cite{Heinemann2019}, which reported a strong northeastward deflection of at least $30^{\circ}$ between 1.3 and 3~$R_s$, attributed to the influence of the open magnetic field configuration associated with a nearby CH, as illustrated in Fig.~\ref{magnetCHsurface}. Additionally, the evolution of the selected magnetic field lines during the first 4 hours of the CME simulation shows that the flux-rope field evolves coherently into the outer corona, consistent with the findings of \cite{Cheng_2014}, further validating our modelling approach.

  \subsection{Evolution of coronal structures during the CME simulation}\label{sec:CMEsimulation}
   In this work, the CME flux rope is located near the Sun–Earth meridian plane, which is approximately perpendicular to the Sun–STEREO-A meridian plane. To further validate the observation-based CME simulation, we compare the evolution of simulated white-light pB images, synthesized in the Sun–Earth meridian plane, with the corresponding pB observations from the STEREO-A spacecraft. We also demonstrate the evolution of the radial velocity, plasma density and temperature at $3~R_s$ during the CME simulation. Additionally, we present the evolution of some physical parameters along the Sun–Earth line and along $30^{\circ}~ {\rm N}$ on the Sun–Earth meridian plane, corresponding to the velocity peak displayed in the panel at $3~R_s$, to further illustrate the impact of the CME on coronal structures.
   
\begin{figure*}[htpb!]
\begin{center}     \includegraphics[width=0.9\linewidth,trim=1 1 1 1, clip]{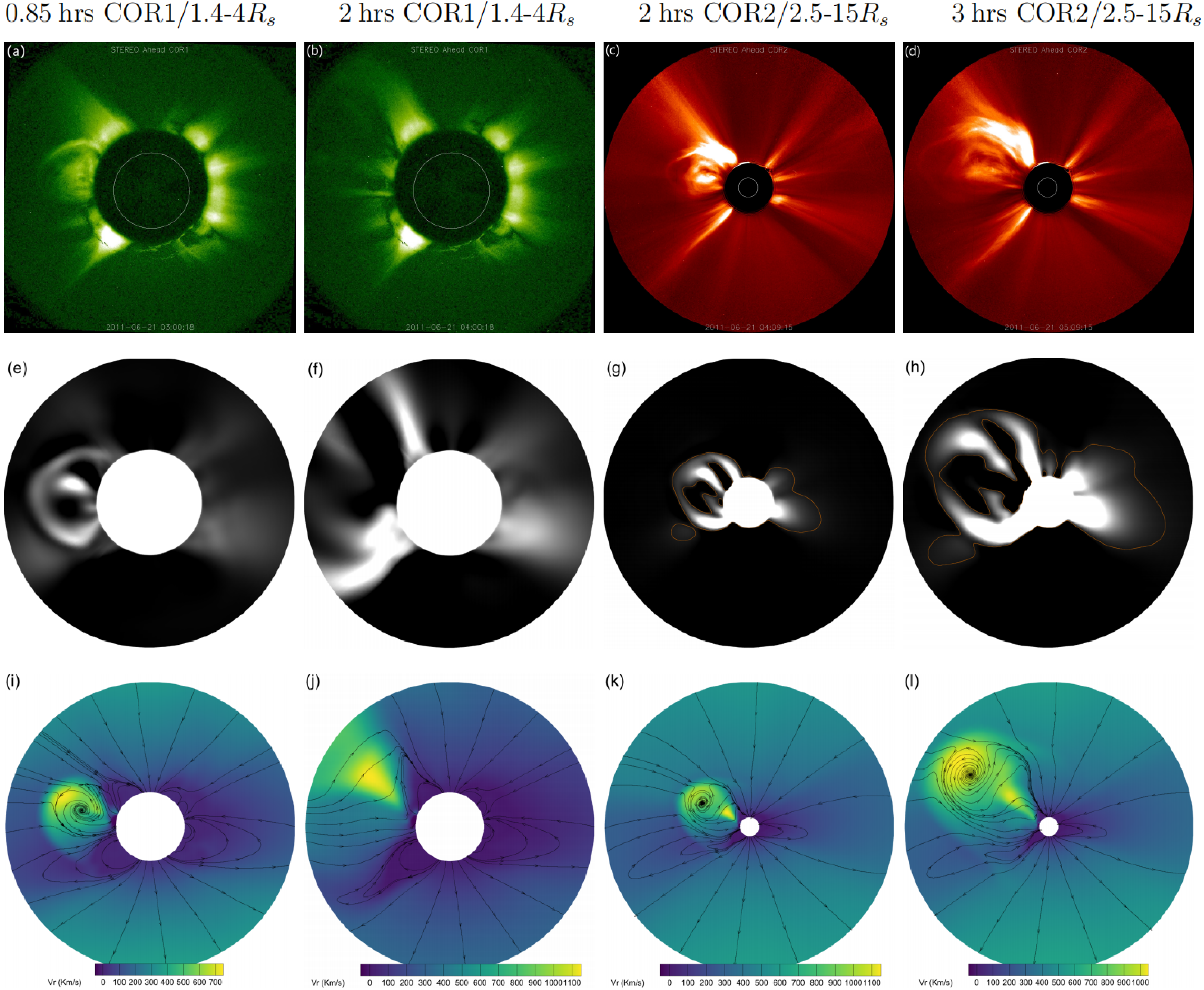}
\end{center}
  \caption{White-light pB images observed from COR1 (a, b) and COR2 (c, d) onboard STEREO-A, corresponding pB images synthesized from simulation results (e-h), along with the simulated radial velocity along the same line of sight as STEREO-A ranging from $1~R_s$ to $4~R_s$ (i, j) and from $1~R_s$ to $15~R_s$ (k, l). The black solid lines overlaid on the radial velocity contours represent selected magnetic field lines.  }\label{pBVrSky}
\end{figure*}
In Fig.~\ref{pBVrSky}, we compare white-light pB images ranging from $1.4~R_s$ to $4~R_s$ (a, b, e, f, i, j) and from $2.5~R_s$ to $15~R_s$ (c,d, g, h, k, l) observed by COR1 (a, b) innermost and COR2 (c, d) outer coronagraph of the SECCHI instrument suite onboard the STEREO-A spacecraft and synthesized from the simulation results calculated by the the MHD model described in Section \ref{sec:Description of the novel MHD Model} (e-h). In these pB images, bright and dark regions represent the high- and low-density coronal structures, respectively. We also show the corresponding radial velocity $V_r \left( {\rm km~ s^{-1}}\right)$ distributions, overlaid with selected magnetic field lines in the same meridional plane ranging from $1~R_s$ to $4~R_s$ (i, j) and from $1~R_s$ to $15~R_s$ (k, l).

It can be seen that the simulated pB images exhibit a typical three-part CME morphology consisting of a bright front, a dark cavity, and a bright core. In the observational data, a clear separation between the bright core and the bright front is evident. However, in the simulation, this separation becomes increasingly blurred in the CME propagation direction at 2 and 3 hours after eruption, although the dark cavity remains pronounced on the lateral sides of the bright core. This discrepancy may be attributed to the decreasing spatial resolution of the grid mesh with increasing heliocentric distance. In addition, the use of oversimplified empirical heating source terms \citep{Wang_2022,wang2025sipifvmtimeevolvingcoronalmodel}, which do not account for the significant structural variations in the corona during the CME evolution, may further cause the deviation from observations.
Moreover, it is interesting to notice that the outlines of the simulated pB images around the CME are generally consistent with the observations from COR1 and COR2 onboard STEREO-A. In particular, the leading edge of the simulated pB closely matches the observed front at 0.85, 2, and 3 hours after the CME onset. This agreement demonstrates that our MHD modelling approach is able to accurately reproduce both the morphological structure and the propagation speed of the observed CME. 

By combining the velocity and 2D magnetic field line distributions (i–l) with the simulated pB images (e–h), we conclude that the bipolar streamer in the southern part of the east limb is displaced northward by approximately $15^{\circ}$ following its interaction with the CME, resulting in improved agreement with the observed pB images. It demonstrates that incorporating transient solar events into coronal simulations can significantly improve the agreement between the simulated large-scale coronal structures and observations. Additionally, the brightest pB feature within the CME core aligns with the center of the spiral-wound CME magnetic field lines, and the corresponding radial velocity is lower than that of the surrounding plasma. This indicates that the simulated high-density structures within the CME flux rope field are also generally associated with relatively low velocities.

\begin{figure*}[htpb!]
\begin{center}     \includegraphics[width=0.9\linewidth,trim=1 1 1 1, clip]{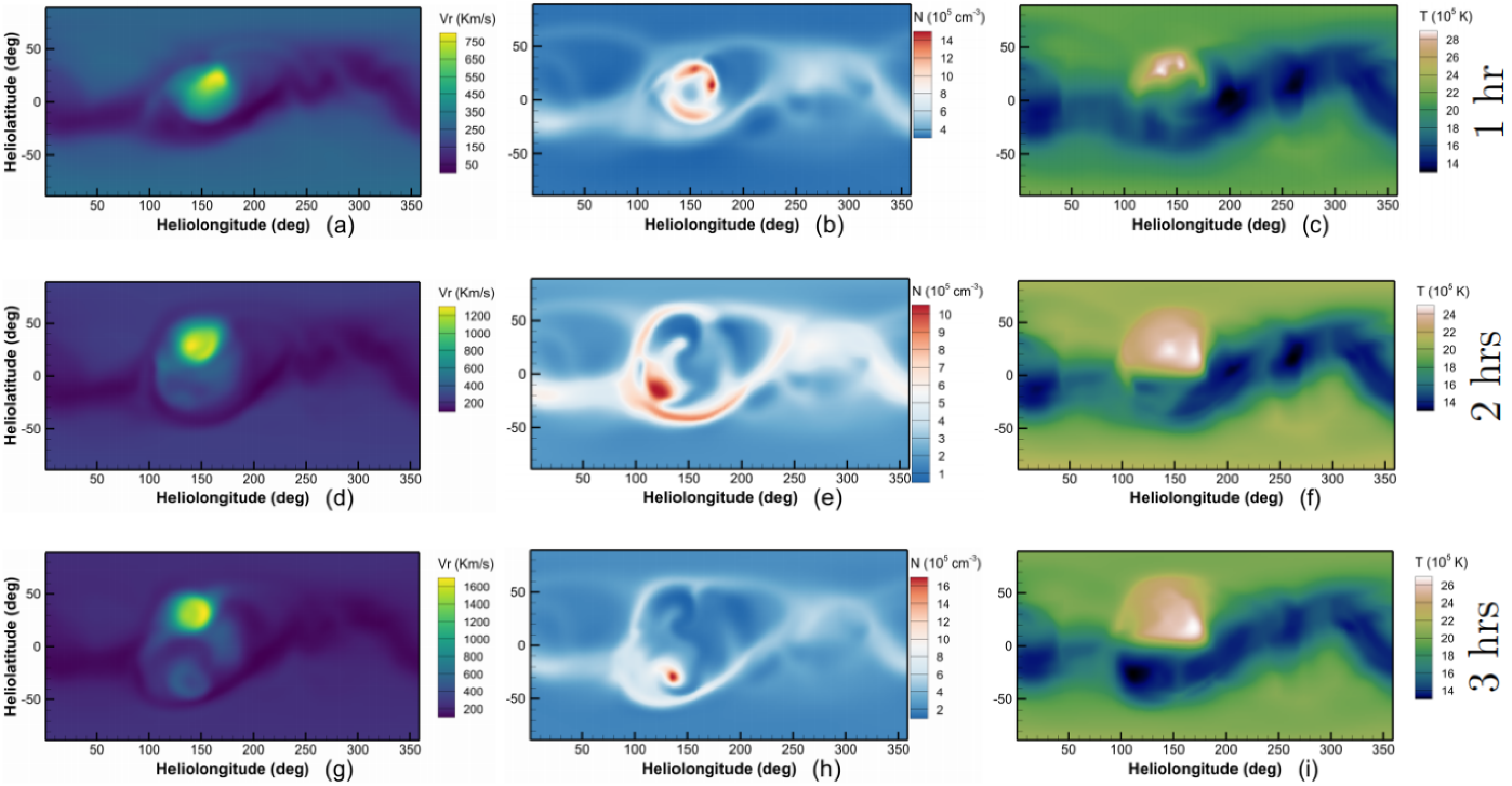}
\end{center}
  \caption{Evolution of radial velocity $V_r$ (a, d, g), in unit of $\rm km~ s^{-1}$, proton number density $N$ (b, e, h), in unite of $10^5~\rm cm^{-3}$, and temperature $T$ (c, f, i), in unite of $10^5~{\rm K}$ at 3 $R_s$. The 1st, 2nd and 3rd rows denote the simulation results at 1, 2 and 3 hours of the CME simulation.}\label{NVrTat3Rs}
\end{figure*}
In Fig.\ref{NVrTat3Rs}, we present the distributions of simulated radial velocity $V_r$ ($\mathrm{km~s^{-1}}$), plasma density $N$ ($10^5~\mathrm{cm^{-3}}$), and temperature $T$ ($10^5~\mathrm{K}$) at $3~R_s$, calculated at 1 (a, b, c), 2 (d, e, f), and 3 (g, h, i) hours of the CME simulation, respectively. 
The regions exhibiting significantly increased velocity, plasma density, and temperature correspond to areas swept by the CME. These regions are encircled by a low-speed, high-density closed band in the background corona at $3~R_s$, which is observed to expand from 1 to 2 hours in the CME simulation due to interaction with the CME. The velocity peak gradually shifts from $(\theta,\phi) = (25^{\circ}\mathrm{N},165^{\circ})$ at 1 hour, with a maximum value of approximately $800~\mathrm{km~s^{-1}}$, to $(\theta,\phi) = (30^{\circ}\mathrm{N},150^{\circ})$ at 2 and 3 hours, reaching maximum values of about $1300~\mathrm{km~s^{-1}}$ and $1700~\mathrm{km~s^{-1}}$, respectively.
Meanwhile, the two plasma density peaks flanking the velocity peak at 1 hour evolve into a distinct '3'-shaped low-density structure centered around $(\theta,\phi) = (0^{\circ},155^{\circ})$ at 2 and 3 hours. New density peaks emerge at $(\theta,\phi) = (20^{\circ}\mathrm{N},120^{\circ})$ and $(\theta,\phi) = (30^{\circ}\mathrm{N},~135^{\circ})$, corresponding to regions of relatively low velocity within the CME-affected area at those times. Correspondingly, the temperature rises significantly from the background value of approximately $1.8 \times 10^5~\mathrm{K}$ to about $2.8 \times 10^5~\mathrm{K}$ near $(\theta,\phi) = (30^{\circ}\mathrm{N},145^{\circ})$ at 1 hour, $2.5 \times 10^5~\mathrm{K}$ near $(\theta,\phi) = (20^{\circ}\mathrm{N},160^{\circ})$ at 2 hours, and $2.7 \times 10^5~\mathrm{K}$ near $(\theta,\phi) = (20^{\circ}\mathrm{N},~165^{\circ})$ at 3 hours.

\begin{figure*}[htpb!]
\begin{center}     \includegraphics[width=0.9\linewidth,trim=1 1 1 1, clip]{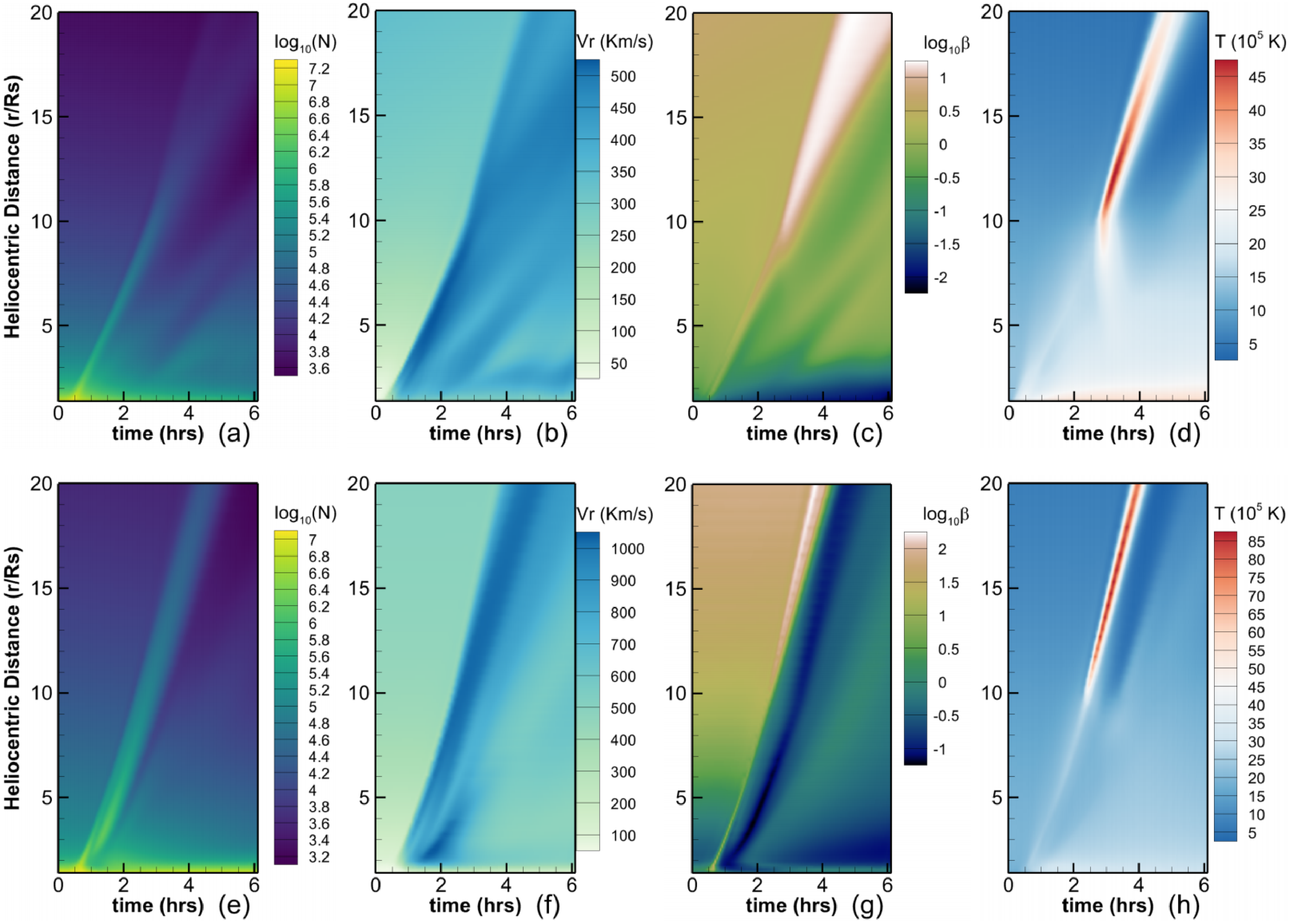}
\end{center}
  \caption{2 D timing diagrams of the base-10 logarithms of the simulated plasma number density $\log_{10}(N)$ (a, e), with N in unite of ${\rm cm}^{-3}$, of the radial velocity $V_r$ (b, f), in unit of $\mathrm{km~s^{-1}}$, of the base-10 logarithms of plasma $\beta$ (c, g), denoted by $\log_{10}(\beta)$, and of the temperature $T$ (d, h), in unite of $10^5~{K}$, along two selected lines ranging from $1.4~R_s$ to $20~R_s$ over the first 6 hours the CME simulation. The first row shows the simulation results along the Sun-to-Earth line, while the second row shows the simulation results along $30^{\circ}~ {\rm N}$ on the Sun–Earth meridian plane.}\label{NVrBetaT1d3to20Rs}
\end{figure*}
In Fig.~\ref{NVrBetaT1d3to20Rs}, we display the 2 D timing diagrams of the base-10 logarithms of the simulated plasma number density $\log_{10}(N)$, radial velocity $V_r \left( {\rm km~ s^{-1}}\right)$, base-10 logarithms of the plasma $\beta$ $\log_{10}(\beta)$, and temperature $T \left(10^5~{\rm K}\right)$ during the first 6 hours of the CME simulation. These 2 D timing diagrams demonstrate the evolution of the simulated parameters ranging from $1.4~R_s$ to $20~R_s$ along the Sun-to-Earth line (a-d), here we denote it as $\rm Line_E$, and along $30^{\circ}~ {\rm N}$ on the Sun–Earth meridian plane, here we denote it as $\rm Line_{30}$. 

It can be seen that approximately 0.25 hours after the CME onset, the plasma density, radial velocity, and temperature along the selected lines begin to increase as the CME approaches, and then decrease as it passes. These CME-induced changes in coronal structure are more pronounced along $\rm Line_{30}$, which is roughly aligned with the CME propagation direction, than along $\rm Line_E$. Notably, for both lines, the increase in plasma density is more evident closer to the Sun, particularly within $10~R_s$, while the temperature increase is more significant beyond $10~R_s$. As the CME passes, both parameters drop to values lower than the previous background levels before gradually recovering to a value comparable to the ambient conditions.
As for the velocity along $\rm Line_{30}$, the CME front experiences acceleration within the first $7~R_s$, reaching a peak and then maintaining a speed of around $1000~\mathrm{km~s^{-1}}$. After the CME departs, the velocity begins to decline but remains higher than the pre-event background speed. 
Along $\rm Line_E$, the velocity peak reaches a maximum of approximately $550~\mathrm{km~s^{-1}}$ between 3 and $9~R_s$, then gradually decreases to about $450~\mathrm{km~s^{-1}}$ and increases the background velocity around this value during the following CME simulation. 
Meanwhile, along $\rm Line_{30}$, the plasma $\beta$ initially shows a slight increase, followed by a rapid decrease to approximately $10^{-1.5}$ as the CME passes, and then gradually rises to about $10^{-0.5}$ after the CME has moved through. Along $\rm Line_E$, $\beta$ also drops below 1 during the CME passage. From these 2 D timing diagrams, we can conclude that our model captures the shock caused by super-Alfv{\'e}nic CME. Additionally, the significant difference in plasma heating within $10~R_s$ compared to beyond $10~R_s$, possibly caused by the manual switching of the heat conduction term from the Spitzer form to the collisionless form around $10~R_s$, highlights the need for a more self-consistent thermodynamic treatment in CME simulations. 

\section{Concluding remarks}\label{sec:conclusion}
   In this paper, we improve the temporal integration accuracy of the recently developed SIP-IFVM coronal model \citep{WangSubmitted,wang2025sipifvmtimeevolvingcoronalmodel} by implementing the ESDIRK2 temporal integration method to advance the solutions in time and adopting the BDF2 pseudo-time marching method to solve the solutions in intermediate stages of ESDIRK2. To improve numerical stability, we design an approximate linearisation technique for the implicitly solved intermediate stages in ESDIRK2. After establishing the MHD coronal model, we use it to reconstruct the background corona and then introduce an observation-based RBSL flux rope, with its position, axis, and magnetic flux derived from observational data, to the background corona to trigger a real CME event. Moreover, the simulation results are compared with observed coronal structures to validate the observation-based CME modelling approach in producing quasi-realist CME evolutions. 
   Furthermore, we reset $\mathbf{B}_1$ to zero after each physical time step in both background coronal and time-dependent CME simulations to validate the extended magnetic field decomposition strategy proposed by \cite{wang2025sipifvmtimeevolvingcoronalmodel}, which introduces a temporally piecewise-constant field to accommodate part of the non-potential field and help keep $\mathbf{B}_1$ sufficiently small throughout the simulation.
   
   Comparison of the extended magnetic field decomposition method with the conventional magnetic field decomposition method in the background coronal simulation shows that both methods produce consistent results, with average relative differences of only $2.21\%$ in proton number density, $1.06\%$ in radial velocity, and $0.92\%$ in magnetic field strength. Moreover, the extended magnetic field decomposition method successfully enables faster-than-real-time CME simulations consistent with observations, whereas the conventional method fails at the initial stage of the CME simulation. This indicates that the extended magnetic field decomposition method not only improves the numerical stability of MHD models in dealing with time-evolving low-$\beta$ issues in the coronal simulations, but also performs effectively in the relatively high-frequent CME simulations involving complex magnetic structures and low-$\beta$ regions.

   Comparison of simulation results with EUV observations from the 193~\AA\ channel of AIA onboard SDO demonstrates that inserting the observation-based flux rope into the background corona improves the consistency of the simulated open-field regions with the observed coronal hole distribution. This highlights the necessity of incorporating high-resolution magnetic field observations into global coronal simulations to more accurately reproduce realistic coronal structures. 
   In addition, the evolution of selected magnetic field lines shows that the simulation successfully reproduces the observed significant northeastward deflection in the CME propagation direction during the first hour. Furthermore, the simulated pB images around the flux rope at 0.85, 2, and 3 hours are consistent with COR1 and COR2 observations from STEREO-A, demonstrating that our MHD model well captures both the morphological structure and propagation speed of the observed CME. Incorporating the CME event into the background corona also notably improves the agreement with observed pB images from COR1/STEREO-A after the CME passes through the region, indicating that including transient solar events in coronal simulations can significantly improve the consistency between simulated large-scale coronal structures and observations. 
   
   These results demonstrate that adopting an observation-based RBSL flux rope in a fully implicit thermodynamic MHD coronal model enables us to perform quasi-realistic faster-than-real-time CME simulations consistent with observations. 
   The implicit method adopted in this model allows for the flexibility of selecting relatively large time steps to improve computational efficiency. The ESDIRK2 scheme, coupled with a BDF2 pseudo-time marching method, improves the temporal accuracy of the CME simulations. Additionally, the approximate linearisation technique applied to the implicitly solved intermediate stages further improves the numerical stability of the ESDIRK2 method. Moreover, the extended magnetic field decomposition helps to address the time-evolving low-$\beta$ issues and improves the numerical stability of the MHD model in CME simulations involving rapidly changing and complex strong magnetic fields.
   Given that CMEs are among the primary drivers of space weather \citep{Vourlidas2019, Zhang2023}, yet there remains a lack of numerically stable, efficient, and operationally applicable modeling approaches for their early propagation and shock evolution in the sub-Alfv{\'e}nic corona \citep{Vourlidas2019}, the novel coronal model and CME simulation approach developed in this research, which helps to address this critical gap, is expected to have the potential to contribute significantly to improving space weather forecasting capabilities.

   Although this observation-based CME simulation performed by a fully implicit thermodynamic MHD coronal model, integrated with the extended magnetic field decomposition strategy, offers significant advantages and is a promising tool for timely and accurate simulations of the realistic CME events involving complex and strong magnetic field in practical space weather forecasting, there is still considerable room for further improvement. 
Synchronized magnetograms \citep{Upton2014a,Downs2025PSI} are required to address the limitation of current synoptic magnetograms, where magnetic fields at different longitudes are observed at different times, for the purpose of reproducing more realistic inner-boundary magnetic field and improving the fidelity of simulated background coronal structures, thereby improving the accuracy of CME simulations. More accurate measurements of the photospheric magnetic fields in the polar regions are also required to reproduce more realistic coronal structures. The time-evolving background coronal modelling during CME simulation is desired to more accurately calculate the interaction between background coronal structures and the CME. Additionally, finer grid meshes and higher-order spatially accurate numerical methods are necessary to better resolve the detailed structures of rapidly evolving CMEs, thereby improving the overall fidelity of the simulations. Accordingly, higher computational efficiency and more advanced numerical algorithms will be essential to support more accurate CME simulations for practical applications, such as daily space weather forecasting.
Moreover, self-consistently simulating the formation and evolution of CMEs by numerical methods, rather than manually inserting a flux rope, is crucial for enabling the reliable prediction of CME events before they emerge in the corona.
Besides, incorporating more realistic thermodynamic mechanisms, such as using physically consistent rather than empirically defined heating source terms, and accounting for field-aligned thermal conduction around CMEs beyond $10~R_s$, where coronal simulations often assume isotropic, collisionless heat conduction, but plasma $\beta$ typically decreases as a CME passes through, may further improve the realism of CME simulations.

There are also several issues worth further investigation, which we plan to explore in future work to improve our numerical models.
 \begin{enumerate}
     \item Advances in the form of more, better, and diversified remote and in-situ observational data are a definite driver of model development in operational space weather predictions. This applies especially but not exclusively for empirical models, since they rely most heavily on observations for model development and application. In contrast, changes in computer architecture seem to be a potential driver with certain lead times. The full exploitation of computing power alone in a heterogeneous CPU${\big/}$GPU cluster \cite[e.g.,][]{Feng_2013Chinese,Feng_2013,WANG2019181} significantly improves the overall performance of MHD-based space weather models. Therefore, in future work, we plan to implement GPU acceleration in our implicit MHD model following grid refinement to enable faster-than-real-time, extremely high-fidelity coronal and CME simulations.   
     
     \item Relevantly, the non-linear force-free field extrapolation initiated MHD simulations are also found to be successful in simulating the dynamics of solar transient events in the solar corona \cite[e.g.,][]{Jiang_2013,Jiang201605}. Using the corresponding simulated MHD coronal results around an active region as an initial CME driver instead of specific flux rope can be promising in realistically modeling solar transient events. Therefore, we plan to use local active region models \citep[e.g.,][]{Jiang201605,Amari2018,Zhong2021} to define the initial conditions for specific regions within the global coronal background. We will then carry out a time-evolving global coronal simulation \citep{W_SubmittedCOCONUT,wang2025COCONUTMayEvent,wang2025sipifvmtimeevolvingcoronalmodel}, with mesh refinement applied around the corresponding active regions, to achieve a more self-consistent CME simulation. This approach is similar to the recent data-driven MHD simulation of a sunspot-rotating active region that led to a solar eruption \citep{Jiang2023}, but in our case, it will be implemented within a thermodynamic MHD coronal model directly driven by time-evolving, high-fidelity observed magnetic fields.

     \item Recently, \citet{Mackay_2022} employed idealized bipoles with self-helicity in a magnetic flux transport model \citep{Mackay_2006,Yeates2013} and the Advective Flux Transport (AFT) model \citep{Upton2014a} to generate a series of radial magnetic field distributions used to drive their global magnetofrictional coronal model. Their study demonstrates that the bipole dataset derived from the AFT model has greater potential for reproducing both highly non-potential and near-potential magnetic field structures showing good agreement with both on-disk and limb observations.
     In our future work, we plan to incorporate the AFT model \citep{Upton2014a}, along with data from the Solar Orbiter Polarimetric and Helioseismic Imager \citep{Loeschl2024}, horizontal velocities inferred from observations using time-distance helioseismology \citep{Zhao2012,Yalim_2017}, and AI-generated data based on STEREO EUV observations \citep{Jeong_2020,Jeong_2022}, to produce a more realistic real-time evolution of the magnetic field at the inner boundary of our MHD coronal model. Also, we may attempt to use an analytical bipole formulation to initialize the energized fields around active regions, simplifying the implementation of self-consistent CME simulations.
     
 \end{enumerate}

\textbf{\textit{Acknowledgments}}: 
The authors thank Prof. Xueshang Feng, Dr. Xiaojing Liu, Dr. Man Zhang, Dr. Yuhao Zhou, Dr. Fan Zhang and Mr. Yuhang Gao for their valuable comments. This project has received funding from the European Research Council Executive Agency (ERCEA) under the ERC-AdG agreement No. 101141362 (Open SESAME). 
These results were also obtained in the framework of the projects FA9550-18-1-0093 (AFOSR), C16/24/010  (C1 project Internal Funds KU Leuven), G0B5823N and G002523N (WEAVE) (FWO-Vlaanderen), and 4000145223 (SIDC Data Exploitation (SIDEX), ESA Prodex).
This work is also supported by the National Natural Science Foundation of China (grant No. 42030204) and the BK21 FOUR program of the Graduate School, Kyung Hee University (GS-1-JO-NON-20242364).
The resources and services used in this work were provided by the VSC (Flemish Supercomputer Centre), funded by the Research Foundation – Flanders (FWO) and the Flemish Government. This work utilises data obtained by the Global Oscillation Network Group (GONG) program, managed by the National Solar Observatory and operated by AURA, Inc., under a cooperative agreement with the National Science Foundation. The data were acquired by instruments operated by the Big Bear Solar Observatory, High Altitude Observatory, Learmonth Solar Observatory, Udaipur Solar Observatory, Instituto de Astrof{\'i}sica de Canarias, and Cerro Tololo Inter-American Observatory. The authors also acknowledge the use of the STEREO/SECCHI data produced by a consortium of the NRL (US), LMSAL (US), NASA/GSFC (US), RAL (UK), UBHAM (UK), MPS (Germany), CSL (Belgium), IOTA (France), and IAS (France).

\appendix
\section{Position of the observation-based axis path of the flux rope}\label{sec:FRaxispathData}
\begin{table}[ht]
\centering
\caption{Position of the observation-based axis path of the flux rope, described in a heliographic coordinate system where the Earth corresponds to $0^\circ$ in longitude.}
\begin{tabular}{rrr}
\toprule
\makecell{\textbf{Longitude}\\(degree, $0^\circ$ at center)} & \makecell{\textbf{Latitude}\\(degree, $0^\circ$ at equator)} & \makecell{\textbf{Heliocentric}\\\textbf{distance} ($R_s$)} \\
\midrule
12.54340  & 14.67630  & 1.02311 \\
11.61810  & 14.45380  & 1.03232 \\
11.53700  & 13.90240  & 1.04809 \\
11.35980  & 13.94890  & 1.05612 \\
11.02740  & 14.08160  & 1.06425 \\
10.66710  & 14.16930  & 1.07018 \\
10.37070  & 14.24720  & 1.07538 \\
10.02630  & 14.33170  & 1.07855 \\
9.73710   & 14.43770  & 1.07987 \\
8.62419   & 14.07660  & 1.08761 \\
7.67697   & 13.27660  & 1.09255 \\
6.99884   & 12.57070  & 1.09066 \\
6.49704   & 12.09390  & 1.08942 \\
5.78038   & 11.23940  & 1.08886 \\
5.10986   & 10.58070  & 1.09122 \\
4.33122   & 9.99759   & 1.09734 \\
3.23655   & 9.30368   & 1.10144 \\
1.48051   & 8.59362   & 1.11432 \\
0.84934   & 8.25828   & 1.12119 \\
-0.04904  & 8.45587   & 1.12084 \\
-0.52731  & 8.49373   & 1.11200 \\
-1.50782  & 8.99747   & 1.09544 \\
-1.88784  & 9.35278   & 1.08999 \\
-2.29134  & 9.81673   & 1.08407 \\
-2.61798  & 10.31190  & 1.07711 \\
-2.85774  & 10.76190  & 1.07225 \\
-3.07082  & 11.29800  & 1.06602 \\
-3.11348  & 11.89480  & 1.05967 \\
-3.05357  & 12.47450  & 1.05243 \\
-2.75490  & 13.00520  & 1.04729 \\
-2.47523  & 13.46370  & 1.03969 \\
-2.13328  & 14.03270  & 1.03398 \\
-1.80474  & 14.50970  & 1.02756 \\
-1.33152  & 15.15540  & 1.01592 \\
-1.08557  & 15.55510  & 1.00574 \\
\bottomrule
\end{tabular}
\end{table}  

\pagebreak

\bibliographystyle{aasjournal}
\bibliography{SIPandCOCONUT}



\end{document}